\documentclass[12pt]{article}

\newcommand{\be}{\begin{equation}}
\newcommand{\ee}{\end{equation}}
\newcommand{\bi}[1]{\vspace{-3mm} \bibitem{#1}}

\usepackage{epsfig,amsmath,amssymb,graphics,graphicx} 

\usepackage{amscd}
\usepackage{pb-diagram}
\usepackage{hyperref}
\usepackage{amsfonts}
\usepackage{epsfig}

\begin{document}

\begin{center}

{\it Annals of Physics. Vol.334. (2013) 1-23.}
\vskip 3mm

{\bf \large Fractional Power-Law Spatial Dispersion in Electrodynamics } \\
\vskip 3mm

\vskip 7mm
{\bf \large Vasily E. Tarasov} \\
\vskip 3mm
{\it Skobeltsyn Institute of Nuclear Physics,\\ 
Lomonosov Moscow State University, 
Moscow 119991, Russia} \\
{E-mail: tarasov@theory.sinp.msu.ru} \\
\vskip 3mm

{\it Departamento de An\'alisis Matem\'atico, \\ 
Universidad de La Laguna, 
38271 La Laguna, Tenerife, Spain.} 

\vskip 7mm
{\bf \large Juan J. Trujillo} \\
\vskip 3mm
{\it Departamento de An\'alisis Matem\'atico, \\ 
Universidad de La Laguna,  
38271 La Laguna, Tenerife, Spain.} \\
{E-mail: jtrujill@ullmat.es} \\

\begin{abstract}
Electric fields in non-local media with power-law spatial dispersion are discussed. 
Equations involving a fractional Laplacian in the Riesz form that describe 
the electric fields in such non-local media are studied. 
The generalizations of Coulomb's law  and Debye's screening for power-law non-local media are characterized. 
We consider simple models with anomalous behavior of plasma-like media with power-law spatial dispersions. 
The suggested fractional differential models for these plasma-like media 
are discussed to describe non-local properties of power-law type. 
\end{abstract}

\end{center}




\newpage

\section{Introduction}

Fractional calculus is dedicated to study the integrals and 
derivatives of any arbitrary real (or complex) order. 
It has a long history from 1695 \cite{Ross,MKM}. 
The first book dedicated specifically to study the theory of
fractional integrals and derivatives and their applications, 
is the book by Oldham and Spanier \cite{OS} published in 1974.
There exists a remarkably comprehensive 
encyclopedic-type monograph by Samko, Kilbas and Marichev \cite{SKM}. 
Many other publications including different fractional models had appeared from 1980. 
For example, see \cite{KST}-\cite{PH11}. 
Fractional calculus and the theory of integro-differential equations of non-integer orders 
are powerful tools to describe the dynamics of anomalous systems and 
processes with power-law non-locality, long-range memory and/or fractal properties.

Spatial dispersion is called the dependence of 
the tensor of the absolute permittivity of the medium on the wave vector \cite{SR,ABR,AR-1}. 
This dependence leads to a number of phenomena, 
such as the rotation of the plane of polarization, anisotropy of cubic crystals 
and other \cite{AG-1,AG-2,AG-3,AR-2,KR,LL-8,Halevi,UFN-1,UFN-2,UFN-3,UFN-4,UFN-5}.
The spatial dispersion is caused by non-local connection 
between the electric induction ${\bf D}$  and the electric field ${\bf E}$. 
Vector ${\bf D}$ at any point ${\bf r}$ of the medium is not uniquely defined by
the values of ${\bf E}$ at this point. It also depends on the values of ${\bf E}$ 
at neighboring points ${\bf r}^{\prime}$, located near the point ${\bf r}$.

Plasma-like medium is medium in which the presence of free charge carriers, creating as they move in the medium, 
electric and magnetic fields, which significantly distorts the external field and the effect on the motion of 
the charges themselves \cite{SR,ABR,AR-1}.  
The term "plasma-like media" refers to media with high spatial dispersion. 
These media are ionized gas, metals and semiconductors, molecular crystals and colloidal electrolytes. 
The term "plasma-like media" was introduced in 1961, by  Viktor P. Silin and Henri A. Rukhadze 
in the book "The electromagnetic properties of the plasma and plasma-like media" \cite{SR}.

In Section 2 the basic concepts and well-known equations of electrodynamics of continuous media are considered to fix the notation. 
In Section 3, we consider power-law type generalizations of  Debye's permittivity  
and generalizations of the correspondent equations for electrostatic potential
by involving the fractional generalization of the Laplacian.
The simplest power-law forms of the longitudinal permittivity 
and correspondent equations for the electrostatic potential are suggested.
The power-law type deformation of Debye's screening and Coulomb's law are discussed.
These suggested simple models allows us to demonstrate new possible types of 
an anomalous behavior of media with fractional power-law type of non-locality. 
In Section 4 the description of weak spatial dispersions of power-law type in the plasma-like media is discussed.
The fractional generalizations of the Taylor series are used for this description. 
The correspondent power-law deformation of Debye's screening 
and Coulomb's law are considered.
A short conclusion is given in Section 5.
In Appendix 1, we suggest a short introduction to the Riesz fractional derivatives and integrals.
In Appendix 2, the fractional Taylor formulas of different types are described.


In this section we review some basic concepts and well-known equations of electrodynamics of continuous media, to fix the notation.  For details see \cite{SR,ABR,AR-1}.

The behavior of electric fields (${\bf E},{\bf D}$), 
magnetic fields (${\bf B}, {\bf H}$), charge density $\rho$, and current density ${\bf j}$ 
is described by the well-known Maxwell's equations 
\be \label{3-3-ME1a}
\operatorname{div} {\bf D}(t,{\bf r})=\rho(t,{\bf r}),
\ee
\be \label{3-3-ME2a}
\operatorname{curl} {\bf E}(t,{\bf r}) = 
-\frac{\partial {\bf B}(t,{\bf r})}{\partial t},
\ee
\be \label{3-3-ME3a}
\operatorname{div} {\bf B}(t,{\bf r})=0,
\ee
\be \label{3-3-ME4a}
\operatorname{curl} {\bf H}(t,{\bf r}) = 
{\bf j}(t,{\bf r})+\frac{\partial {\bf D}(t,{\bf r})}{\partial t} .
\ee

The densities $\rho(t,{\bf r})$ and ${\bf j}(t,{\bf r})$ 
describe an external source of field. 
We assume that the external sources of electromagnetic field are given.
The vector ${\bf E}(t,{\bf r})$ is the electric field strength and
the vector ${\bf D}(t,{\bf r})$ is the electric displacement field.
In free space, the electric displacement field is equivalent to flux density.
The field ${\bf B}(t,{\bf r})$ is the magnetic induction and 	
the vector ${\bf H}(t,{\bf r})$ is the magnetic field strength.

In the case of the linear electrodynamics the constitutive equations (material equations) are linear relations.
For electromagnetic fields which are changed slowly in the space-time, 
we have the constitutive equations (material equations) in the well-known form
\be \label{LDE}
D_i(t,{\bf r}) = \varepsilon_{ij} E_j(t,{\bf r}) ,
\ee
\be \label{LBH}
B_i(t,{\bf r}) = \mu_{ij} H_j(t,{\bf r}) ,
\ee
where $\varepsilon_{ij}$ and $\mu_{ij}$ are second-rank tensors.
For fields varying in space rapidly, we should consider the influence of the field at remote points ${\bf r}^{\prime}$ on the electromagnetic properties of the medium at a given point ${\bf r}$.
The field at a given point ${\bf r}$ of the medium will be determined not only the value of the field at this point, but  the field in the areas of environment, where the influence of the field is transferred.  For example, it can be caused by the transport processes in the medium.
Therefore, we should use non-local space relations instead of equations (\ref{LDE}), (\ref{LBH}). These non-local relations take into account space dispersion. For linear electrodynamics we have 
the following relation between electric fields ${\bf E}$ and ${\bf D}$ given by
\be \label{NLDE}
D_i(t,{\bf r}) = \int_{\mathbb{R}^3} \hat \varepsilon_{ij}({\bf r},{\bf r}^{\prime}) \, 
E_j(t,{\bf r}^{\prime}) \, d {\bf r}^{\prime},
\ee
and
\be \label{NLDE-b}
B_i(t,{\bf r}) = \int_{\mathbb{R}^3} \hat \mu_{ij}({\bf r},{\bf r}^{\prime}) \, 
H_j(t,{\bf r}^{\prime}) \, d {\bf r}^{\prime} .
\ee

If the medium is not limited in space and homogeneous, then the kernel of the integral operator is a function of the position difference 
${\bf r}-{\bf r}^{\prime}$,
\be \label{NLDE2}
D_i(t,{\bf r}) = \int_{\mathbb{R}^3}  
\hat \varepsilon_{ij}({\bf r}-{\bf r}^{\prime}) \, E_j(t,{\bf r}^{\prime}) \, d {\bf r}^{\prime} .
\ee

In this case we can use the Fourier transform. 
The direct and inverse Fourier transforms ${\cal F}$ and ${\cal F}^{-1}$, for suitable functions, are given by 
\be
({\cal F} f) ({\bf k})= {\cal F}[f({\bf r})]({\bf k})= 
\int_{\mathbb{R}^3} e^{ - i ({\bf k}{\bf r})} \, f({\bf r}) \, d^3 {\bf r},
\ee
\be
\hat g({\bf r})=({\cal F}^{-1} g) ({\bf r})= {\cal F}^{-1}[ g({\bf k})]({\bf r})= \frac{1}{(2 \pi)^3}
\int_{\mathbb{R}^3} e^{ + i ({\bf k}{\bf r})} \, g({\bf k}) \, d^3 {\bf k}.
\ee
We does not use the hat for ${\bf D}(t,{\bf r})$ and ${\bf E}(t,{\bf r})$ to have usual notation. From the context, it will be easy to understand if the field is considered in the space-time of its the Fourier transforms.

Then electric field will be represented
as a set of plane monochromatic waves, for which space-time dependence are defined by the function $exp\{i {\bf k} {\bf r}- i \omega t\}$. Therefore, 
relation (\ref{NLDE2}) has the form
\be \label{NLDE3}
D_i(\omega,{\bf k}) = \varepsilon_{ij}({\bf k}) E_j (\omega,{\bf k}).
\ee

The function $\varepsilon_{ij}({\bf k})$ is called the tensor 
of the absolute permittivity of the material:
\be
\varepsilon_{ij}({\bf k}) = \int_{\mathbb{R}^3}   e^{ -i {\bf k} {\bf r}}  \,
\hat \varepsilon_{ij}({\bf r}) \, d {\bf r}^{\prime} .
\ee

Even for an isotropic linear medium, the dependence of the tensor  $\varepsilon_{ij}({\bf k})$
of the wave vector ${\bf k}$ preserves tensor form 
\cite{SR,ABR,AR-1}.
In this case we have
\be
\varepsilon_{ij}({\bf k})= \left(\delta_{ij} - \frac{k_ik_j}{|{\bf k}|^2} \right) \,
\varepsilon_{\perp} (|{\bf k}|)+
\frac{k_ik_j}{|{\bf k}|^2} \, \varepsilon_{\parallel}  (|{\bf k}|) ,
\ee
where $\varepsilon_{\perp}(|{\bf k}|)$ - the transverse permittivity,
and $\varepsilon_{\parallel}  (|{\bf k}|)$ - the longitudinal permittivity.

The Maxwell's equations for the electromagnetic fields have the well-known form  \cite{SR,ABR,AR-1}
\be \label{FME1}
i ({\bf k},{\bf E}(\omega,{\bf k})) \, \varepsilon ({\bf k})= \rho (\omega,{\bf k}) ,
\ee
\be \label{FME2}
[{\bf k},{\bf E}(\omega,{\bf k})] = \omega \, {\bf B} (\omega,{\bf k}) ,
\ee
\be \label{FME3}
({\bf k},{\bf B}(\omega,{\bf k})) =0,
\ee
\be \label{FME4}
\frac{i}{\mu (\bf k)} \, [{\bf k},{\bf B}(\omega,{\bf k})] = 
-i \omega \varepsilon ({\bf k}) \, {\bf E} (\omega,{\bf k}) 
+ {\bf j} (\omega,{\bf k}) .
\ee

In these equations we have neglected the frequency dispersion.
This can be done when the inhomogeneous field can be approximately regarded as static.

In the case of a static external field sources in the environment 
can create a inhomogeneous electric field ${\bf E} (t,{\bf r}) ={\bf E} ({\bf r})$. The electric field in the medium is given by
\be \label{Pot-E}
{\bf E} ({\bf r}) = - \operatorname{grad} \Phi ({\bf r}) , 
\ee
where $\Phi ({\bf r})$ is a scalar potential of electric field.
Relation (\ref{Pot-E}), applying the Fourier transform, can be written by
\be \label{E=kP}
{\bf E}({\bf k}) = - i {\bf k} \, \Phi_{\bf k}.
\ee
Therefore, substituting (\ref{E=kP}) into (\ref{FME1}), we obtain
\be \label{FME1b}
|{\bf k}|^2 \, \varepsilon_{\parallel} \, (|{\bf k}|) \, \Phi_{\bf k} = \rho_{\bf k} ,
\ee
where $\rho_{\bf k}=\rho (0,{\bf k})$. Note that equation (\ref{FME1b}) does not   depend of the transverse permittivity $\varepsilon_{\perp}(|{\bf k}|)$.

When the field source in the medium is the resting point charge, then 
the charge density is described by delta-distribution
\be \label{delta}
\rho({\bf r}) = Q \, \delta^{(3)} ({\bf r}).
\ee
Therefore the electrostatic potential of the point charge in the isotropic medium, according to the equation (\ref{FME1b}), has the form
\be
\Phi ({\bf r}) = \frac{Q}{(2 \pi)^3} 
\int_{\mathbb{R}^3} e^{ + i {\bf k}({\bf r} }) \, 
\frac{1}{|{\bf k}|^2 \, \varepsilon_{\parallel} (|{\bf k}|)} \, d^3 {\bf k}, 
\ee
where $\Phi ({\bf r})$ is the electric potential created by a point charge $Q$ 
at a distance $|{\bf r}|$ from the charge.

Let us note the well-known case \cite{SR,ABR,AR-1} is such that
$\varepsilon_{\parallel}  (|{\bf k}|) = \varepsilon_0$,
where the constant $\varepsilon_0$ is the vacuum permittivity 
($\varepsilon_0 \approx 8.854 \, 10^{-12} F \cdot m^{-1}$). 
Substituting $\varepsilon_{\parallel}  (|{\bf k}|) = \varepsilon_0$ into  (\ref{FME1b}), we obtain
\be \label{FDEB-C}
|{\bf k}|^{2} \, \Phi_{\bf k} = \frac{1}{\varepsilon_0} \rho_{\bf k}.
\ee
The inverse Fourier transform of (\ref{FDEB-C}) gives
\be \label{Eq-2}
\Delta \Phi ({\bf r}) = - \frac{1}{\varepsilon_0} \rho({\bf r}) ,
\ee
where $\Delta =  {\partial^2}/{\partial x^2}+  {\partial^2}/{\partial y^2}+  {\partial^2}/{\partial z^2}$
is the 3-dimensional Laplacian, and
\be
{\cal F}[ \Delta f({\bf r})]({\bf k})= 
- |{\bf k}|^2 \, {\cal F}[ f({\bf r})]({\bf k}) =
- |{\bf k}|^2 \, {\hat f}({\bf k}) ,
\ee
where ${\hat f}({\bf k}) = {\cal F}[ f({\bf r})]({\bf k})$.
As a result, the electrostatic potential of the point charge (\ref{delta}) has Coulomb's form
\be \label{Pot-C}
\Phi ({\bf r}) = \frac{1}{4 \pi \varepsilon_0} \, \frac{Q}{|{\bf r}|} .
\ee

The second well-known case \cite{SR,ABR,AR-1} is such that
\be \label{Debye1}
\varepsilon_{\parallel} (|{\bf k}|) = \varepsilon_0 \Bigl(1+\frac{1}{r^2_D |{\bf k}|^2} \Bigr) . 
\ee
Substituting (\ref{Debye1}) into (\ref{FME1b}), we obtain
\be \label{FDEB-D}
\Bigl( |{\bf k}|^2 + \frac{1}{r^2_D} \Bigr) \, \Phi_{\bf k}  
= \frac{1}{\varepsilon_0} \rho_{\bf k} .
\ee
Then, using the inverse Fourier transform of (\ref{FDEB-C}), we get
\be \label{Eq-2b}
\Delta \Phi ({\bf r}) - \frac{1}{r^2_D} \Phi ({\bf r}) = - \frac{1}{\varepsilon_0} \rho({\bf r}) .
\ee
As a result, we have the screened potential of the point charge (\ref{delta}) in Debye's form:
\be \label{Pot-D}
\Phi ({\bf r}) = \frac{1}{4 \pi \varepsilon_0} \frac{Q}{|{\bf r}|} \, 
\exp \Bigl( - \frac{|{\bf r}|}{r_D} \Bigr) , 
\ee
where $r_D$ is Debye's radius of screening. It is easy to see that Debye's potential differs from Coulomb's potential by factor $C_D(|{\bf r}|) = \exp (-|{\bf r}|/r_D)$. Such factor is a decay factor for Coulomb's law,
where the parameter $r_D$ defines the distance over which significant charge separation can occur. Therefore, Debye's sphere is a region with Debye's radius $r_D$, in which there is an influence of charges, and outside of which charges are screened.


\section{Fractional Power-Law of Non-Locality and Generalized Debye's Screening}

In this section, we consider power-law type generalizations 
of Debye's permittivity (\ref{Debye1}), 
and generalizations of the correspondent equations for electrostatic potential $\Phi ({\bf r})$ of the form
\be \label{Eq-2c}
\Delta \Phi ({\bf r}) - \frac{1}{r^2_D} \Phi ({\bf r}) = - \frac{1}{\varepsilon_0} \rho({\bf r}) 
\ee
by involving the fractional generalization of the Laplacian \cite{Riesz-1,Riesz-2,SKM,KST}.


\subsection{A Power-law Generalizations}

In this section we consider the simplest power-law forms of the longitudinal permittivity 
$\varepsilon_{\parallel} (|{\bf k}|)$ and correspondent equations for the electrostatic potential. The suggested simple models allows us to consider
new possible types of an anomalous behavior of media with fractional power-law type of non-locality. We introduce some deformation of power-law type to well-known model of Debye's screening.

The generalized model is described by the deformation of two terms in equation (\ref{Debye1}) for permittivity in the from
\be  \label{peps-3}
\varepsilon_{\parallel} (|{\bf k}|) = \varepsilon_0 
\Bigl( |{\bf k}|^{\alpha-2} + \frac{1}{r^2_D \, |{\bf k}|^{2-\beta}} \Bigr) . 
\ee
The parameter $\alpha$ characterizes the deviation from Coulomb's law due to non-local properties of the medium.
The parameter $\beta$ characterizes the deviation from Debye's screening due to non-integer power-law type of non-locality in the medium.

Substituting (\ref{peps-3}) into (\ref{FME1b}), we obtain
\be \label{FDEB-3}
\Bigl( |{\bf k}|^{\alpha} + \frac{1}{r^2_D} |{\bf k}|^{\beta} \Bigr) \, \Phi_{\bf k} 
= \frac{1}{\varepsilon_0} \rho_{\bf k},
\ee
and using the inverse Fourier transform of (\ref{FDEB-3}), we have
\be \label{Eq-3}
((-\Delta)^{\alpha/2} \Phi) ({\bf r}) + \frac{1}{r^2_D} ((-\Delta)^{\beta/2} \Phi)({\bf r}) 
= \frac{1}{\varepsilon_0} \rho({\bf r}),
\ee
where $(-\Delta)^{\alpha/2}$ and  $(-\Delta)^{\beta/2}$ are the Riesz fractional Laplacian, see, for instance, \cite{Riesz-1,Riesz-2,SKM,KST} and Appendix 1. 
Note that ${\bf r}$ and $r_D$ are dimensionless variables.

In order to describe the properties of two types of deviations separately, we consider the following special cases of the proposed model.

\vspace{3mm} \noindent 1) Fractional model of non-local deformation of Coulomb's law in the media with spatial dispersion defined by equation (\ref{peps-3}) with $\beta=0$, given by
\be \label{peps-1}
\varepsilon_{\parallel} (|{\bf k}|) = \varepsilon_0 
\Bigl(|{\bf k}|^{\alpha-2} + \frac{1}{r^2_D |{\bf k}|^2} \Bigr) . 
\ee
Then equation (\ref{FME1b}) has the form 
\be \label{FDEB-1}
\Bigl( |{\bf k}|^{\alpha} + \frac{1}{r^2_D} \Bigr) \, \Phi_{\bf k} 
= \frac{1}{\varepsilon_0} \rho_{\bf k} ,
\ee
and the equation for electrostatic potential is
\be \label{Eq-1}
((-\Delta)^{\alpha/2} \Phi)({\bf r}) + \frac{1}{r^2_D} \Phi ({\bf r}) 
= \frac{1}{\varepsilon_0} \rho({\bf r}) .
\ee
This model allows us to describe a possible deviation from  Coulomb's law in the media with nonlocal properties defined by power-law type of spatial dispersion.

\vspace{3mm} \noindent 2) Fractional model of non-local deformation of Debye's screening in the media with spatial dispersion is defined by equation (\ref{peps-3}) with $\alpha=2$, is given by
\be \label{peps-2}
\varepsilon_{\parallel} (|{\bf k}|) = \varepsilon_0 
\Bigl( 1 + \frac{1}{r^2_D \, |{\bf k}|^{2- \beta}} \Bigr) . 
\ee
Equation (\ref{FME1b})  with (\ref{peps-2}) lead to
\be \label{FDEB-2}
\Bigl( |{\bf k}|^{2} + \frac{1}{r^2_D} |{\bf k}|^{\beta} \Bigr) \, \Phi_{\bf k} 
= \frac{1}{\varepsilon_0} \rho_{\bf k},
\ee
and then the corresponding equation for generalized potential is given by 
\be \label{Eq-2d}
-\Delta \Phi({\bf r}) + \frac{1}{r^2_D} ((-\Delta)^{\beta/2} \Phi)({\bf r})
=  \frac{1}{\varepsilon_0} \rho({\bf r}),
\ee
which involve two different differential operators.
Such model allows us to describe a possible deviation from 
Debye's screening by non-local properties of the plasma-like media with the generalized power-law type of spatial dispersion. \\

The behavior of electrostatic potentials for fractional differential models described by equations (\ref{Eq-1}) and (\ref{Eq-2d}) will be consider in Section 3.3. To the mentioned model can be find a explicit solution in terms of a Green type function. Also we will describe analytic solutions of the fractional differential equations (\ref{Eq-3}).


\subsection{General Fractional Power-Law Type of Non-Locality}

In the more general case, we can consider the following power-law form
\be  \label{peps-G}
\varepsilon_{\parallel} (|{\bf k}|) = \varepsilon_0 
\Bigl( \sum^m_{j=1} a_j |{\bf k}|^{\alpha_j-2} + \frac{a_0}{|{\bf k}|^2} \Bigr). 
\ee
Substitution of (\ref{peps-G}) into (\ref{FME1b}), and using the inverse Fourier transform gives the fractional partial differential equation
\be \label{FPDE-1}
\sum^m_{j=1} a_j ((-\Delta)^{\alpha_j/2} \Phi) ({\bf r}) + a_0 \Phi ({\bf r})= \frac{1}{\varepsilon_0} \rho({\bf r}),
\ee
where $\alpha_m > ... > \alpha_1>0$, and $a_j \in \mathbb{R}$ 
($1 \leq j \leq m$) are constants.

We apply the Fourier method to solve fractional equation (\ref{FPDE-1}), 
which is based on the relation 
\be \label{FFL}
{\cal F}[ (-\Delta)^{\alpha/2} f({\bf r})]({\bf k})= 
|{\bf k}|^{\alpha} \, \hat f({\bf k}).
\ee
Applying the Fourier transform ${\cal F}$ to both sides of (\ref{FPDE-1}) and using (\ref{FFL}), we have
\be
({\cal F} \Phi)({\bf k}) = \frac{1}{\varepsilon_0} \left( \sum^m_{j=1} a_j |{\bf k}|^{\alpha_j}+a_0 \right)^{-1} 
({\cal F} \rho)({\bf k}) .
\ee

The fractional analog of the Green function (see Section 5.5.1. in \cite{KST}) is given by
\be \label{FGF}
G_{\alpha}({\bf r})= {\cal F}^{-1} \Bigl[ \left( \sum^m_{j=1} a_j |{\bf k}|^{\alpha_j}+
a_0 \right)^{-1} \Bigr] ({\bf r})=
\int_{\mathbb{R}^3} \left( \sum^m_{j=1} a_j |{\bf k}|^{\alpha_j}+a_0 \right)^{-1} \
e^{ + i ({\bf k},{\bf r}) } \, d^3 {\bf k} ,
\ee
where $\alpha=(\alpha_1,...,\alpha_m)$.

The following relation
\be \label{3-1}
\int_{\mathbb{R}^n} e^{  i ({\bf k},{\bf r}) } \, f(|{\bf k}|) \, d^n {\bf k}= 
\frac{(2 \pi)^{n/2}}{ |{\bf r}|^{(n-2)/2}} 
\int^{\infty}_0 f( \lambda) \, \lambda^{n/2} \, J_{n/2-1}(\lambda |{\bf r}|) \, d \lambda
\ee
holds (see Lemma 25.1 of \cite{SKM}) for any suitable function $f$
such that the integral in the right-hand side of (\ref{3-1}) is convergent. 
Here $J_{\nu}$ is the Bessel function of the first kind. As a result, the Fourier transform of a radial function is also a radial function.

On the other hand, using (\ref{3-1}), the Green function (\ref{FGF}) can be represented (see Theorem 5.22 in \cite{KST})
in the form of the one-dimensional integral involving the Bessel function $J_{1/2}$ of the first kind 
\be \label{G-1}
G_{\alpha} ({\bf r}) =\frac{|{\bf r}|^{-1/2}}{(2 \pi)^{3/2}} 
\int^{\infty}_0 \left( \sum^m_{j=1} a_j |\lambda|^{\alpha_j}+a_0 \right)^{-1} 
\lambda^{3/2} \, J_{1/2} (\lambda |{\bf r}|) \, d \lambda,
\ee
where we use $n=3$ and $\alpha=(\alpha_1,...,\alpha_m)$.
Note that 
\be
J_{1/2} (z) = \sqrt{\frac{2}{\pi z}} \, \sin (z).
\ee

If $\alpha_m > 1$ and $A_m \ne 0$, $A_0 \ne 0$, then equation (\ref{FPDE-1})  (see, for example, Section 5.5.1. pages 341-344 in  \cite{KST}) has a particular solution is given by (\ref{phi-G}). Such particular solution is represented in the form of the convolution of the functions $G({\bf r})$ and $\rho({\bf r})$ as follow
\be \label{phi-G}
\Phi({\bf r})= \frac{1}{\varepsilon_0} \int_{\mathbb{R}^3} G_{\alpha} ({\bf r} - {\bf r}^{\prime}) \, 
\rho ({\bf r}^{\prime}) \, d^3 {\bf r}^{\prime},
\ee
where the Green function $G_{\alpha}(z)$ is given by (\ref{G-1}).


Therefore, we can consider the fractional partial differential equation (\ref{FPDE-1}) with $a_0=0$ and $a_1 \ne 0$, when $m \in \mathbb{N}$, $m \ge 1$, and also the case where $\alpha_1< 3$, $\alpha_m > 1$, $m \ge 1$, $a_1 \ne 0$, $a_m \ne 0$, $\alpha_m > ... > \alpha_1>0$, which is given by
\be \label{FPDE-3}
\sum^m_{j=1} a_j ((-\Delta)^{\alpha_j/2} \Phi) ({\bf r}) = \frac{1}{\varepsilon_0} \rho({\bf r}) .
\ee
The above equation has the following particular solution (see Theorem 5.23 in \cite{KST}), given by
\be \label{phi-G3}
\Phi({\bf r})= \frac{1}{\varepsilon_0} \int_{\mathbb{R}^3} G_{\alpha} ({\bf r} - {\bf r}^{\prime}) \, 
\rho ({\bf r}^{\prime}) \, d^3 {\bf r}^{\prime} ,
\ee
with
\be \label{G-3}
G_{\alpha} ({\bf r}) =\frac{|{\bf r}|^{-1/2}}{(2 \pi)^{3/2}} 
\int^{\infty}_0 \left( \sum^m_{j=1} a_j |\lambda|^{\alpha_j} \right)^{-1} 
\lambda^{3/2} \, J_{1/2} (\lambda |{\bf r}|) \, d \lambda .
\ee
These particular solutions allows us to describe
electrostatic field in the plasma-like media with 
the spatial dispersion of power-law type.


\subsection{Potentials for Particular Cases of Non-Integer Power-Law Type Non-Locality}

In this section we will study the properties of electrostatic potentials for fractional differential models mentioned in Section 3.1. 

Here we consider particular solutions of the following fractional partial differential equation
\be \label{FPDE-4}
((-\Delta)^{\alpha/2} \Phi) ({\bf r}) + a_{\beta} ((-\Delta)^{\beta/2} \Phi) ({\bf r}) = \frac{1}{\varepsilon_0} \rho({\bf r}) ,
\ee
where $1<\alpha$, $0<\beta < \alpha$, $\beta <3$, and $a_{\beta}=r^{-2}_D$.  Note that ${\bf r}$ and $r_D$ are dimensionless. Equation (\ref{FPDE-4}) is the fractional partial differential equation (\ref{FPDE-3}) with $m=1$, and such equation has the following particular solution 
\be \label{phi-G4}
\Phi({\bf r})= \frac{1}{\varepsilon_0} \int_{\mathbb{R}^3} 
G_{\alpha, \beta} ({\bf r} - {\bf r}^{\prime}) \, 
\rho ({\bf r}^{\prime}) \, d^3 {\bf r}^{\prime},
\ee
where the Green type function is given by
\be \label{G-4}
G_{\alpha, \beta} ({\bf r}) =\frac{|{\bf r}|^{-1/2}}{(2 \pi)^{3/2}} 
\int^{\infty}_0 \left( |\lambda|^{\alpha}+ a_{\beta} |\lambda|^{\beta} \right)^{-1} 
\lambda^{3/2} \, J_{1/2} (\lambda |{\bf r}|) \, d \lambda.
\ee
Therefore, the electrostatic potential of the point charge (\ref{delta}) has form:
\be \label{Pot-2}
\Phi ({\bf r}) = \frac{1}{4 \pi \varepsilon_0} \frac{Q}{|{\bf r}|} \
\cdot C_{\alpha, \beta} (|{\bf r}|) ,
\ee
with
\be \label{Cab1}
C_{\alpha, \beta} (|{\bf r}|) = \frac{2}{\pi} 
\int^{\infty}_0 \frac{ \lambda \, \sin (\lambda |{\bf r}|)}{ 
|\lambda|^{\alpha}+ a_{\beta} |\lambda|^{\beta}  } \, d \lambda,
\ee
where $C_{\alpha, \beta} (|{\bf r}|)$ describes the difference 
of Coulomb's potential.

Now we will study three special cases: 
1) $\alpha \ne 2$, $\alpha >1$ and $\beta=0$; 
2) $\alpha=2$ and $0<\beta<2$; 
3) $\alpha \ne 2$ and $\beta>0$.


\subsubsection{Non-local deformation of Coulomb's law (the case $\beta=0$)}

Fractional model of non-local deformation of Coulomb's law in the media with spatial dispersion is defined by equation (\ref{FPDE-4}) with $\beta=0$, is given by
\be \label{FPDE-2}
((-\Delta)^{\alpha/2} \Phi)({\bf r}) + \frac{1}{r^2_D} \Phi ({\bf r}) 
= \frac{1}{\varepsilon_0} \rho({\bf r}) ,
\ee
where $\alpha>1$, and $a_0=r^{-2}_D>0$.
Using (\ref{Pot-2}) and (\ref{Cab1}), it is easy to see that the electrostatic potential $\Phi ({\bf r})$ differs from 
Coulomb's potential by the factor
\be \label{C-1}
C_{\alpha,0} (|{\bf r}|) =
\frac{2}{\pi} \, 
\int^{\infty}_0 \frac{ \lambda \, \sin (\lambda |{\bf r}|)}{|\lambda|^{\alpha}+a_0 } \, d \lambda.
\ee
Note that Debye's potential differs from Coulomb's potential by the exponential factor $C_D(|{\bf r}|) =\exp (-|{\bf r}|/r_D)$. \\


In Figure 1 we present plots of Debye exponential factor $C_D(|{\bf r}|) =\exp (-|{\bf r}|/r_D)$  and generalized factor $C_{\alpha,0}(|{\bf r}|)$ for 
the different orders of $1.5 < \alpha < 3.0$ and $a_0=r^{-2}_D=1$.\\



\begin{figure}[H]
\begin{minipage}[h]{0.47\linewidth}
\resizebox{8cm}{!}{\includegraphics[angle=-90]{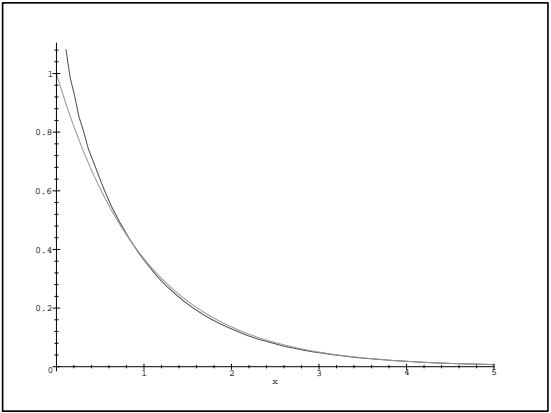}} 
a) \\
\end{minipage}
\hfill
\begin{minipage}[h]{0.47\linewidth}
\resizebox{8cm}{!}{\includegraphics[angle=-90]{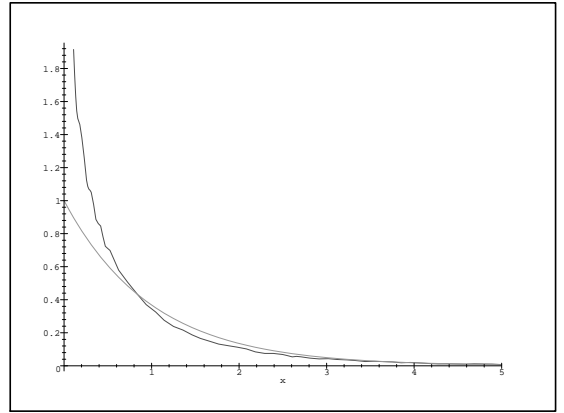}} 
b) \\
\end{minipage}
\vfill
\begin{minipage}[h]{0.47\linewidth}
\resizebox{8cm}{!}{\includegraphics[angle=-90]{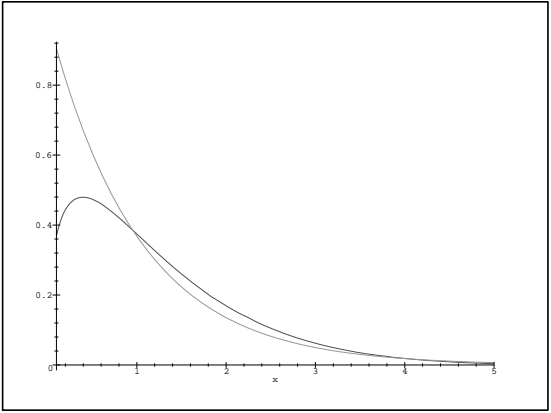}} 
c) \\
\end{minipage}
\hfill
\begin{minipage}[h]{0.47\linewidth}
\resizebox{8cm}{!}{\includegraphics[angle=-90]{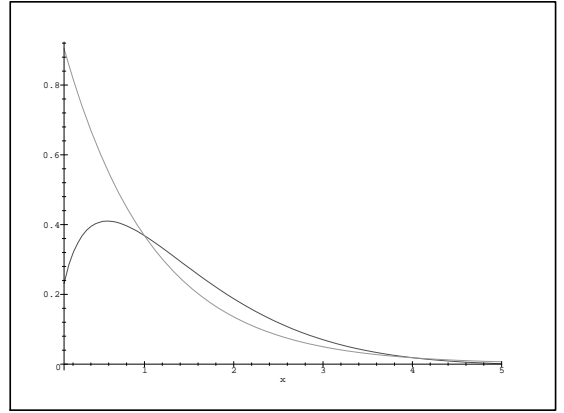}} 
d) \\
\end{minipage}
\caption{Plots of Debye exponential factor $C_D(x) =\exp (-x/r_D)$ with $r_D=1$ and the factors $y=C_{\alpha,0}(x)$ with $a_0=r^{-2}_D=1$ for the orders: a) $\alpha=1.9$, b) $\alpha=1.6$, c) $\alpha=2.5$, d) $\alpha=2.8$. Here $x=|{\bf r}|$ and we use $0<x<5$.}
\label{Plot1}
\end{figure}


Using (Section 2.12.1 No.3. page 169. in \cite{PBM}), we obtain the asymptotic ($|{\bf r}| \to 0$) in the form
\be \label{Ca0-1}
C_{\alpha,0} (|{\bf r}|) \ \approx \ 
\frac{2^{2-\alpha} \, \Gamma((3-\alpha)/2)}{\sqrt{\pi} \, \Gamma(\alpha/2)} 
\frac{1}{|{\bf r}|^{2-\alpha}} , \quad (1<\alpha<2) ,
\ee
\be \label{Ca0-2}
C_{\alpha,0}  (|{\bf r}|) \ \approx \ 
\frac{2^{2-\alpha} \, \Gamma((3-\alpha)/2)}{\sqrt{\pi} \, \Gamma(\alpha/2)} 
|{\bf r}|^{\alpha-2} , \quad (2<\alpha<3) ,
\ee
\be
C_{\alpha,0}  (|{\bf r}|) \ \approx \  
\frac{2 \Gamma(3/ \alpha) \, \Gamma(1-3/\alpha)}{\pi \, \alpha \, a^{1-3/\alpha}_0}
\, |{\bf r}| , \quad (\alpha>3).
\ee
Note that asymptotic (\ref{Ca0-1}-\ref{Ca0-2}) for $1<\alpha<2$ and $2<\alpha<3$ does not depend on the parameter $a_0$.  
We point out that for $\alpha=2$, using (Equation (11) of Section 1.2. in the book \cite{BE}), we obtain Debye's exponents 
$C_{2,0}(|{\bf r}|)=C_D(|{\bf r}|)$.

As a result, the electrostatic potential of the point charge in a media with this type of spatial dispersion will have the form
\be
\Phi ({\bf r}) \ \approx \ \frac{Q}{4 \pi \varepsilon_0} 
\frac{2^{2-\alpha} \, \Gamma((3-\alpha)/2)}{\sqrt{\pi} \, 
\Gamma(\alpha/2)} \,
\frac{1}{|{\bf r}|^{3-\alpha}}
\quad (1<\alpha<2, \quad 2<\alpha<3) 
\ee
on small distances $|{\bf r}| \ll 1$. In the case $\alpha>3$, we have the constant value of the potential for $|{\bf r}| \ll 1$ given by
\be
\Phi ({\bf r})\ \approx \ \frac{1}{4 \pi \varepsilon_0} \frac{Q}{R_{eff}} , \quad (\alpha >3),
\ee
where $R_{eff}$ is an effective sphere radius that is equal to
\be
R_{eff}=\frac{\pi \, \alpha \, a^{1-3/\alpha}_0}{2 \Gamma(3/ \alpha) \, \Gamma(1-3/\alpha)}.
\ee
Therefore, the electric field ${\bf E}$ is equal to zero at small distances $|{\bf r}| \ll 1$. It is well-known that the electric field inside a charged conducting sphere is zero, and that the potential remains constant at the value it reaches at the surface. Then the electric field of a point charge in the media with power-law of spatial dispersion with $\alpha>3$ is analogous to the field inside a conducting charged sphere of the radius $R_{eff}$, for small distances $|{\bf r}| \ll 1$.

The study of the asymptotic behavior of $C_{\alpha, 2}(|{\bf r}|)$ for 
$ |{\bf r}| \to \infty$ is an open question, although we have evidences to can  suggest, as a conjecture, that its asymptotic behavior follow a power-law type also. Also from the corresponding plots, we can observe that the 
$C_{\alpha, 2}(|{\bf r}|)$ decreases more slowly than 
Debye's exponent $C_D(|{\bf r}|)$.

It is easy to proof that $C_{\alpha, 2}(|{\bf r}|)$  has a maximum for the case $2<\alpha<3$ and the maximum does not exists for $1<\alpha<2$, while for the particular case $\alpha=2$ it is well-known that it is the classical exponential Debye's screening.


\subsubsection{Non-local deformation of Debye's screening   (the case $\alpha=2$)}

Fractional model of non-local deformation of Debye's screening in the media with spatial dispersion is described by equation (\ref{FPDE-4}) with $\alpha=2$, given by 
\be \label{FPDE-4-2b}
-\Delta \Phi ({\bf r}) + a_{\beta} ((-\Delta)^{\beta/2} \Phi) ({\bf r}) 
= \frac{1}{\varepsilon_0} \rho({\bf r}) ,
\ee
where $0<\beta<2$. 
The electrostatic potential of the point charge (\ref{delta}) has the following form
\be \label{Pot-2-2b}
\Phi ({\bf r}) = \frac{1}{4 \pi \varepsilon_0} \frac{Q}{|{\bf r}|} \,
\cdot \, C_{2, \beta} (|{\bf r}|) ,
\ee
where the function
\be
C_{2, \beta} (|{\bf r}|) = \frac{2}{\pi} 
\int^{\infty}_0 \frac{ \lambda \, \sin (\lambda |{\bf r}|)}{ 
|\lambda|^2+ a_{\beta} |\lambda|^{\beta}  } \, d \lambda ,
\ee
where $C_{2, \beta} (|{\bf r}|)$ describes the difference of 
 Coulomb's potential

In Figure 2 we present some plots of Debye exponential factor $C_D(|{\bf r}|) =\exp (-|{\bf r}|/r_D)$ and factor $C_{2,\beta}(|{\bf r}|)$ for 
 different orders of $1.5 < \beta < 2$ and $a_{\beta}=1$. \\



\begin{figure}[H]
\begin{minipage}[h]{0.47\linewidth}
\resizebox{8cm}{!}{\includegraphics[angle=-90]{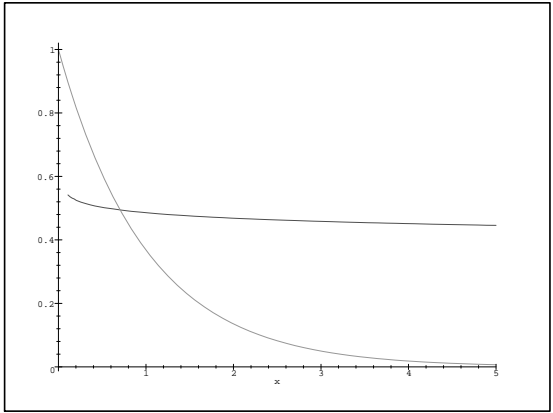}} 
a) \\
\end{minipage}
\hfill
\begin{minipage}[h]{0.47\linewidth}
\resizebox{8cm}{!}{\includegraphics[angle=-90]{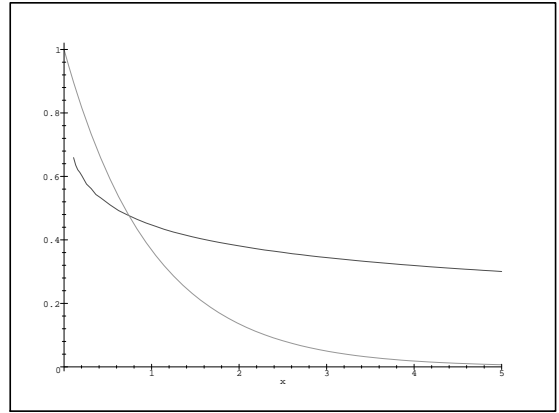}} 
 b) \\
\end{minipage}
\vfill
\begin{minipage}[h]{0.47\linewidth}
\resizebox{8cm}{!}{\includegraphics[angle=-90]{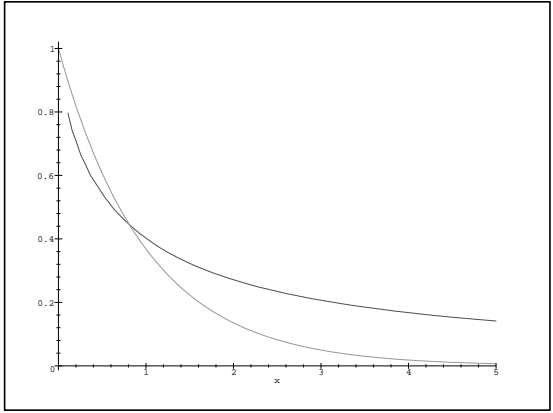}} 
c) \\
\end{minipage}
\hfill
\begin{minipage}[h]{0.47\linewidth}
\resizebox{8cm}{!}{\includegraphics[angle=-90]{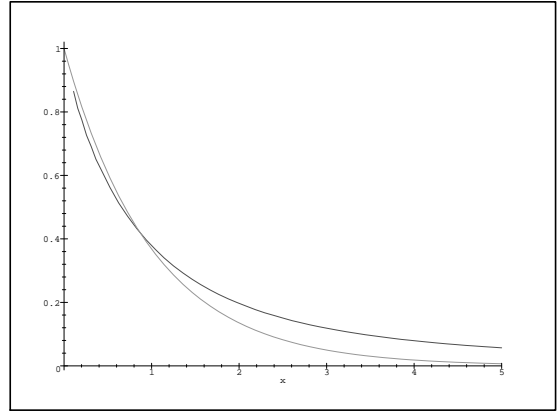}} 
d) \\
\end{minipage}
\caption{Plots of Debye exponential factor $C_D(x) =\exp (-x/r_D)$ with $r_D=1$ and the
factors $y=C_{2,\beta}(x)$ with $a_{\beta}=1$ for the orders: a) $\beta=1.9$, b) $\beta=1.6$, c) $\beta=1.1$, d) $\beta=0.6$. Here $x=|{\bf r}|$ and we use $0<x<5$.}
\label{Plot2}
\end{figure}

%
%

Using (Equation (1) of Section 2.3 in the book \cite{BE}), we obtain the following asymptotic behavior for $C_{2, \beta}(|{\bf r}|)$ with $\beta<2$, when $|{\bf r}| \to \infty$
\be
C_{2,\beta} (|{\bf r}|) = \frac{2}{\pi}
\int^{\infty}_0 \frac{\lambda \sin (\lambda |{\bf r}|)}{\lambda^2+ a_{\beta} \lambda^{\beta}}  \, d \lambda \approx
A_0(\beta) \frac{1}{|{\bf r}|^{2-\beta}} + \sum^{\infty}_{k=1} A_k(\beta) \frac{1}{|{\bf r}|^{(2-\beta)(k+1)}} ,
\ee
where 
\be
A_0(\beta)= \frac{2}{\pi a_{\beta}} \, \Gamma(2-\beta) \, \sin \left( \frac{\pi}{2}\beta \right) ,
\ee
\be
A_k (\beta) = -\frac{2}{\pi a^{k+1}_{\beta}} \int^{\infty}_0  z^{(2-\beta)(k+1)-1} \, \sin(z) \, dz .
\ee

As a result, we have that generalized non-local properties deforms Debye's screening such that the exponential decay is replaced by the following generalized power-law 
\be
C_{2,\beta} (|{\bf r}|) \ \approx \  \frac{A_0}{|{\bf r}|^{2-\beta}} 
\quad (0< \beta<2).
\ee
On the other hand, the electrostatic potential of the point charge in the media with this type of spatial dispersion is given by
\be
\Phi ({\bf r}) \ \approx \ \frac{A_0}{4 \pi \varepsilon_0} \,
\cdot \, \frac{Q}{|{\bf r}|^{3-\beta}} \quad (0< \beta<2)
\ee
on the long distance $|{\bf r}| \gg 1$. \\


\subsubsection{Non-local deformation of Coulomb's law and Debye's screening  (the case $\alpha \ne 2$ and $\beta >0$) together}

The electrostatic potential for non-local  fractional differential model that is described by equation
(\ref{FPDE-4}) includes two parameters $(\alpha, \beta)$,
where $\alpha>\beta > 0$.
In such model non-local properties deforms Coulomb's law and 
Debye's screening such that we have the following fractional power-law decay  
\be
C_{\alpha,\beta} (|{\bf r}|) \ \approx \  \frac{2 \Gamma(2-\beta) \sin(\pi \beta/2)}{ \pi a_{\beta}} \,
\cdot \,  \frac{1}{|{\bf r}|^{2-\beta}} \quad (|{\bf r}| \to \infty), 
\ee
for $0<\beta<3$ and $\alpha> \beta$. Note that this asymptotic behavior $ |{\bf r}| \to \infty$ does not depend on the parameter $\alpha$.
The field on the long distances is determined only by term with $(-\Delta)^{\beta/2}$ ($\alpha>\beta$) that can be interpreted  as a non-local deformation of Debye's (second) term in equation (\ref{Eq-2c}). 

The new type of behavior of the spatial-dispersion media with power-law non-locality is presented by power-law decreasing of the field at long distances instead of exponential decay.

The asymptotic behavior $C_{\alpha,\beta} (|{\bf r}|)$ for $|{\bf r}| \to 0$  is given by
\be \label{Cab-1}
C_{\alpha,\beta} (|{\bf r}|) \ \approx \ 
\frac{2^{2-\alpha} \, \Gamma((3-\alpha)/2)}{\sqrt{\pi} \, \Gamma(\alpha/2)} \,
\cdot \, 
\frac{1}{|{\bf r}|^{2-\alpha}} , \quad (1<\alpha<2),
\ee
\be \label{Cab-2}
C_{\alpha,\beta} (|{\bf r}|) \ \approx \ 
\frac{2^{2-\alpha} \, \Gamma((3-\alpha)/2)}{\sqrt{\pi} \, \Gamma(\alpha/2)} \,
\cdot \,  |{\bf r}|^{\alpha-2} , \quad (2<\alpha<3),
\ee
\be \label{Cab-3}
C_{\alpha,\beta}  (|{\bf r}|) \ \approx \ 
\frac{2}{\alpha \, a^{1-3/\alpha}_{\beta} \, \sin (3 \pi / \alpha)}
\,
\cdot \,  |{\bf r}| , \quad (\alpha>3),
\ee
where we use Euler's reflection formula for Gamma function.
Note that the above asymptotic behavior does not depend on the parameter $\beta$, and relations (\ref{Cab-1}-\ref{Cab-2}) does not depend on $a_{\beta}$.
The field on the short distances is determined only by term with $(-\Delta)^{\alpha/2}$ ($\alpha>\beta$) that can be considered as a non-local deformation of Coulomb's (first) term in equation (\ref{Eq-2c}).

On the other hand, it is remarkable that exist a maximum for the factor $C_{\alpha,\beta} (|{\bf r}|)$ in the case $0<\beta < 2 < \alpha$.


\begin{figure}[H]
\resizebox{15cm}{!}{\includegraphics[angle=0]{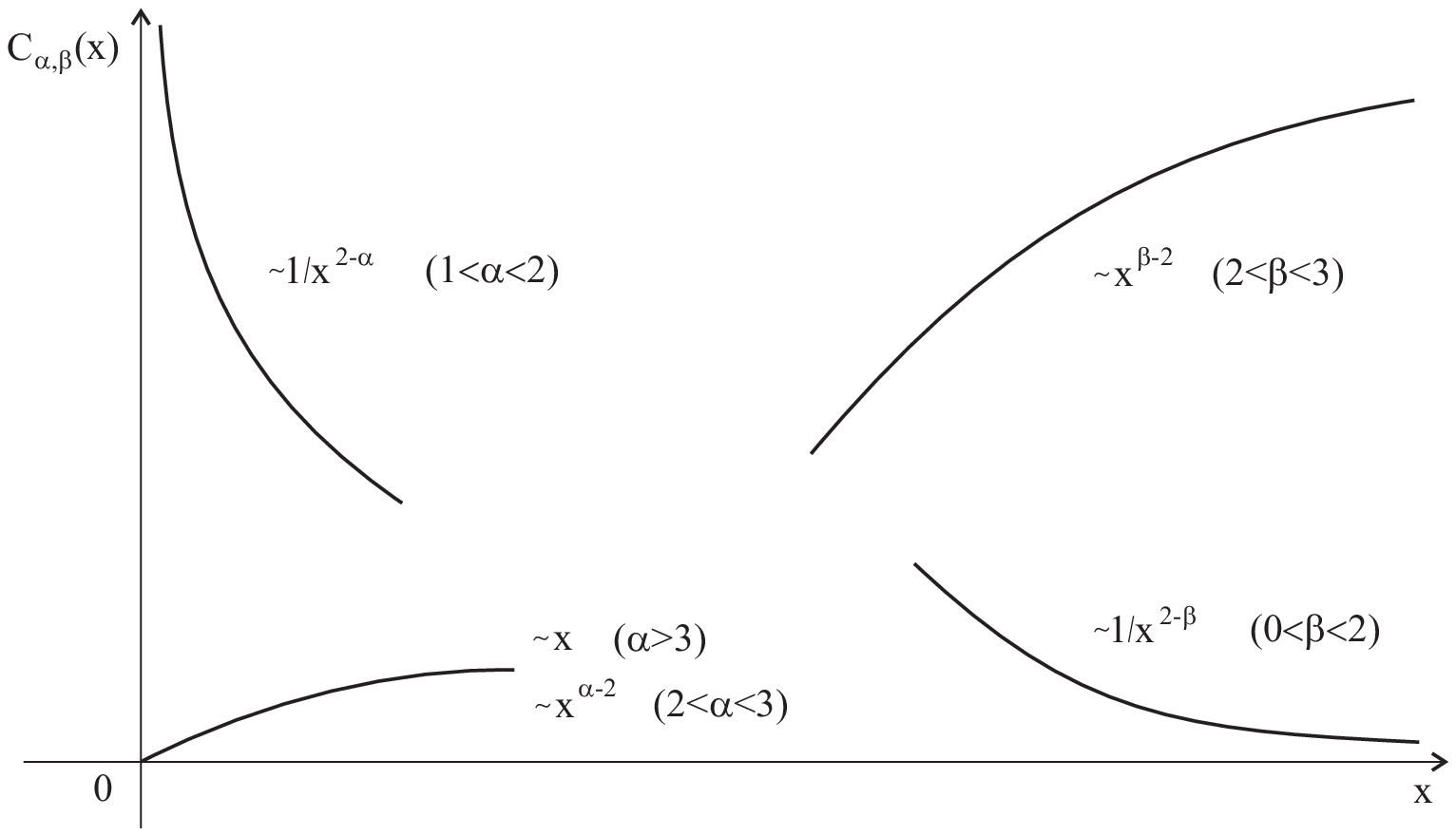}} 
\caption{Plots of general asymptotic behaviors of the factor $y=C_{\alpha,\beta}(x)$.}
\label{Plot-Truji}
\end{figure}

\section{Fractional Weak Spatial Dispersion}

Here we introduce a generalization of well-known weak spatial dispersion for the power-law type of non-locality of the media 
\cite{SR,ABR,AR-1,AG-1,AG-2,AG-3}.

\subsection{Weak Spatial Dispersion}

Let us give an short description of weak spatial dispersion in the plasma-like media (for details see, for instance, \cite{SR,ABR,AR-1}).

Spatial dispersion in electrodynamics is called to the dependence of 
the tensor of the absolute permittivity of the medium on the wave vector \cite{SR,ABR,AR-1}. It is well-known that this dependence leads to a number of phenomena, 
for example the rotation of the plane of polarization, anisotropy of cubic crystals and other \cite{AG-1,AG-2,AG-3,AR-2,KR,LL-8,Halevi,UFN-1,UFN-2,UFN-3,UFN-4,UFN-5}.

The spatial dispersion is caused by non-local connection 
between the electric induction ${\bf D}$  and the electric field ${\bf E}$. 
Vector ${\bf D}$ at any point ${\bf r}$ of the medium is not uniquely defined by
the values of ${\bf E}$ at this point. It also depends on the values of ${\bf E}$ 
at neighboring points ${\bf r}^{\prime}$, located near the point ${\bf r}$.

Non-local connection between ${\bf D}$ and ${\bf E}$ can be understood 
on the basis of qualitative analysis of a simple model of the crystal.
In this model the particles of the crystal lattice (atoms, molecules, ions) 
oscillate about their equilibrium positions and interact with each other.
The equations of oscillations of the crystal lattice particles
with the local (nearest-neighbor) interaction   
gives the partial differential equation of integer orders
in the continuous limits \cite{2006-1,2006-2}.
Note that non-local (long-range) interactions in the crystal lattice
in the continuous limit can give 
a fractional partial differential equations \cite{2006-1,2006-2}.
It was shown in  \cite{2006-1,2006-2} that the equations of oscillations of crystal lattice with long-range interaction are mapped 
into the continuum equation with the Riesz fractional derivative. 

The electric field of the light wave moves charges from their equilibrium positions 
at a given point ${\bf r}$, which causes an additional shift of the charges 
in neighboring and more distant points ${\bf r}^{\prime}$ in some neighborhood.
Therefore, the polarization of the medium, and hence the field ${\bf D}$ depend 
on the values of the electric fields ${\bf E}$ not only in a selected point, 
but also in its neighborhood.
This applies not only to the crystals, but also 
to isotropic media consisting of asymmetric molecules
and plasma-like media \cite{SR,ABR,AR-1}.

The size of the area in which the kernel 
$\hat \varepsilon_{ij} ({\bf r})$ of integral equation (\ref{NLDE3}) 
is significantly determined by the characteristic lengths of interaction $R_0$. 
For different media these lengths can vary widely.
The size of the area of the mutual influence $R_0$ are usually on the order 
of the lattice constant or the size of the molecules (for dielectric media). 
Wavelength of light $\lambda$ is several orders larger than the size of this region, 
so for a region of size $R_0$ value of the electromagnetic field of 
light wave does not change.
By other words, in the dielectric media for optical wavelength $\lambda$ usually holds 
$k R_0 \sim R_0 / \lambda \sim 10^{-3} \ll 1$. 
In such media the spatial dispersion is weak \cite{SR,ABR,AG-1,UFN-5}. 
To analyze it is enough to know the dependence of the tensor 
$\varepsilon_{ij} ({\bf k})$ only for small values 
${\bf k}$ and we can replace the function by the Taylor polynomial
\be \label{Tay1}
\varepsilon_{ij}({\bf k}) = \varepsilon_{ij} +
\gamma_{ijl} k_l + \delta_{ijlm} k_l k_m + ... \quad .
\ee
Here we neglect the frequency dispersion, and so the tensors 
$\varepsilon_{ij}$, $\gamma_{ijl}$, $\delta_{ijlm}$ 
do not depend on the frequency $\omega$.

The tensors in (\ref{Tay1}) are simplified for crystals with high symmetry \cite{AG-1}. 
For an isotropic linear medium, we can use 
\be \label{Tay2}
\varepsilon_{\parallel}  (|{\bf k}|) = \varepsilon +
\gamma |{\bf k}| + \delta |{\bf k}|^2 + ... \quad .
\ee

In order to explain the natural optical activity (for example, optical rotation, gyrotropy) 
is sufficient to consider the linear dependence on ${\bf k}$ in (\ref{Tay1}) and (\ref{Tay2}). 
For non-gyrotropic crystals it is necessary 
to take into account the terms quadratic in ${\bf k}$.

For power-like type of non-locality we should use fractional generalizations 
of the Taylor formula (see Appendix 2).


\subsection{Fractional Taylor series approach}

The weak spatial dispersion in the media with
power-law type of non-locality cannot be describes by
the usual Taylor approximation. The fractional Taylor series is very useful for approximating non-integer power-law functions \cite{WM}. 
To illustrate this point, we consider the non-linear power-law function
\be \label{ve}
\varepsilon_{\parallel}  (|{\bf k}|) = a_{\alpha} |{\bf k}|^{\alpha} +a_0 .
\ee
If we use the usual Taylor series for the function (\ref{ve}) then we have infinite series.

For fractional Taylor formula of Caputo type (see Appendix 2), we need the following known property of the fractional Caputo derivative $_a^CD^{\alpha}_k$  (see, for instance, \cite{KST})
\be \label{T2}
_a^CD^{\alpha}_k (k-a)^{\beta} = \frac{\Gamma(\beta+1)}{\Gamma(\beta-\alpha+1)} \, (k-a)^{\beta-\alpha} , 
\quad (x>a, \ \alpha>0, \ \beta >0) ,
\ee
where $k=|\bf{k}|$. In particular, if $\beta=\alpha$, then
\be
\, _a^CD^{\alpha}_k (k-a)^{\alpha} = \Gamma(\alpha+1), \quad (\, _a^CD^{\alpha}_k )^n (k-a)^{\alpha} =0 . 
\ee

Therefore 
\[ (\, ^CD^{\alpha} \varepsilon_{\parallel} )(0) = \Gamma(\alpha+1), \]
while, the higher order Caputo fractional derivatives of $\varepsilon_{\parallel}  (|{\bf k}|)$, given in (\ref{ve}), are all zero. 
Hence, the fractional Taylor series approximation of such function is exact.
Note that the order of non-linearity of $\varepsilon_{\parallel}  (|{\bf k}|)$ is equal to 
the order of the Taylor series approximation.

\subsection{Weak spatial dispersion of power-law types}

We consider such properties of the media with weak spatial dispersion that is described by the non-integer power-law type of functions 
$\varepsilon_{\parallel}  (|{\bf k}|)$. The fractional differential model is used to describe a new possible type of behavior of complex media with power-law 
non-locality. 

The weak spatial dispersion (and the permittivity) will be called $\alpha$-type,  
if the function $\varepsilon_{\parallel}  (|{\bf k}|)$ satisfies the condition
\be \label{ea1}
\lim_{|{\bf k}| \to 0} 
\frac{\varepsilon_{\parallel}  (|{\bf k}|) - \varepsilon_{\parallel}  (0) }{ \varepsilon_0 \, |{\bf k}|^{\alpha}} =a_{\alpha},
\ee
where $\alpha>0$ and $0<|a_{\alpha}|< \infty$.
Here the constant $\varepsilon_0$ is the vacuum permittivity 
($\varepsilon_0 \approx 8.854 \, 10^{-12} F \cdot m^{-1}$). 

The weak spatial dispersion (the permittivity) will be called $(\alpha,\beta)$-type,  
if the function $\varepsilon_{\parallel}  (|{\bf k}|)$ satisfies the conditions (\ref{ea1}) and
\be \label{ea2}
\lim_{|{\bf k}| \to 0} 
\frac{\varepsilon_{\parallel}  (|{\bf k}|) - \varepsilon_{\parallel}  (0) - 
a_{\alpha} \varepsilon_0 \, |{\bf k}|^{\alpha} }{ \varepsilon_0 |{\bf k}|^{\beta}} = a_{\beta},
\ee
where $\beta>\alpha>0$ and $0<|a_{\beta}|< \infty$.

Note that these definitions are similar to definitions of 
non-local alpha-interactions between particles of 
crystal lattice (see Section 8.6 in \cite{PH7} and \cite{2006-1,2006-2}) 
that give continuous medium equations 
with fractional derivatives with respect to coordinates.

For the weak spatial dispersion of the $(\alpha,\beta)$-type,   
the permittivity can be represented in the form
\be
\varepsilon_{\parallel}  (|{\bf k}|)= \varepsilon_0 (\varepsilon + a_{\alpha} |{\bf k}|^{\alpha} + 
a_{\beta} |{\bf k}|^{\beta}) + R_{\alpha,\beta}(|{\bf k}|) ,
\ee
where $\varepsilon = \varepsilon_{\parallel} (0) / \varepsilon_0$ can be considered 
as the relative permittivity of material, and
\be 
\lim_{|{\bf k}| \to 0} 
\frac{ R_{\alpha,\beta}(|{\bf k}|)}{|{\bf k}|^{\beta}} =0.
\ee
As a result, we can use the following approximation for
weak spatial dispersion 
\be \label{approx-1}
\varepsilon_{\parallel}  (|{\bf k}|)/ \varepsilon_0  \approx \varepsilon + a_{\alpha} |{\bf k}|^{\alpha} + 
a_{\beta} |{\bf k}|^{\beta} .
\ee

If $\alpha=1$ and $\beta=2$, we can use the usual Taylor formula. In this case we have the well-known case of the weak spatial dispersion \cite{AG-1,AG-2,AG-3,AR-2,KR,LL-8,Halevi,UFN-1,UFN-2,UFN-3,UFN-4,UFN-5}
In general, we should use a fractional generalization of the Taylor series (see Appendix 2). If the orders of the fractional Taylor series approximation 
will be correlated with the type of weak spatial dispersion, then the fractional Taylor series approximation of $\varepsilon_{\parallel}  (|{\bf k}|)$ will be exact. 
In the general case, $\beta \ne \alpha$, where $0<\beta- \alpha<1$, we can use the fractional Taylor formula in the Dzherbashyan-Nersesian form (see Appendix 2). For the special cases $\beta=2\alpha$, where $\alpha<1$ and/or $\beta=\alpha+1$, we could use other kind of the fractional Taylor formulas.
In the fractional cases new types of physical effects may exist.

\subsection{Fractional differential equation for electrostatic potential}

We can consider a weak spatial dispersion of the power-law $(\alpha,\beta)$-type. Then substituting (\ref{approx-1}) into (\ref{FME1b}), we obtain
\be \label{Apr-1}
\Bigl( \varepsilon |{\bf k}|^2 + a_{\alpha} |{\bf k}|^{\alpha+2} + 
a_{\beta} |{\bf k}|^{\beta+2} \Bigr) \, \Phi_{\bf k} = \frac{1}{ \varepsilon_0} \rho_{\bf k} ,
\ee
where $\varepsilon =\varepsilon_{\parallel}  (0) \varepsilon_0$ and $\beta > \alpha>0$. 
The inverse Fourier transform of (\ref{Apr-1}) gives
\be \label{Apr-2}
a_{\beta} ( (-\Delta)^{(\beta+2)/2} \Phi)({\bf r}) + a_{\alpha} ( (-\Delta)^{(\alpha+2)/2} \Phi)({\bf r}) - 
\varepsilon \Delta \Phi ({\bf r}) = \frac{1}{ \varepsilon_0} \rho ({\bf r}) .
\ee
This fractional differential equation describes a weak spatial dispersion of the $(\alpha,\beta)$-type.


Equation (\ref{Apr-2}) has the following particular solution 
\be \label{phi-Apr-2}
\Phi({\bf r})= \frac{1}{\varepsilon_0} \int_{\mathbb{R}^3} 
G_{2,\alpha, \beta} ({\bf r} - {\bf r}^{\prime}) \, 
\rho ({\bf r}^{\prime}) \, d^3 {\bf r}^{\prime} ,
\ee
where $G_{2,\alpha, \beta}$ is the Green function of the form
\be \label{G-Apr-2}
G_{2,\alpha, \beta} ({\bf r}) =\frac{|{\bf r}|^{-1/2}}{(2 \pi)^{3/2}} 
\int^{\infty}_0 \left( a_{\alpha} |\lambda|^{\alpha+2}+ 
a_{\beta} |\lambda|^{\beta+2} + \varepsilon |\lambda|^2 \right)^{-1} 
\lambda^{3/2} \, J_{1/2} (\lambda |{\bf r}|) \, d \lambda .
\ee

Therefore, the electrostatic potential of the point charge (\ref{delta}) for this case is given by
\be \label{Pot-Apr-2}
\Phi ({\bf r}) = \frac{1}{4 \pi \varepsilon_0} \frac{Q}{|{\bf r}|} 
 \cdot C_{2, \alpha,\beta} (|{\bf r}|) ,  \ee
where $0<\alpha < \beta$, and the function
\be
C_{2, \alpha,\beta} (|{\bf r}|) =
\frac{2}{\pi} \, \int^{\infty}_0 
\frac{ \lambda \, \sin (\lambda |{\bf r}|)}{ a_{\alpha} |\lambda|^{\alpha+2}+ 
a_{\beta} |\lambda|^{\beta+2} + \varepsilon |\lambda|^2 } \, d \lambda 
\ee
describes the difference between such generalized potential and 
Coulomb's potential. \\


For the weak spatial dispersion of the $\alpha$-type, we have  
\be \label{Apr-3}
a_{\alpha} ( (-\Delta)^{(\alpha+2)/2} \Phi)({\bf r}) - \varepsilon \Delta \Phi ({\bf r}) 
= \frac{1}{ \varepsilon_0} \rho({\bf r}) .
\ee
This equation is a special case of equation (\ref{Apr-2}), where $a_{\beta}=0$. Then, the electrostatic potential of the point charge has form
\be \label{Pot-Apr-3}
\Phi ({\bf r}) = \frac{1}{4 \pi \varepsilon_0} \frac{Q}{|{\bf r}|} \
 \cdot C_{2, \alpha,0}  (|{\bf r}|) , 
\ee
where $0<\alpha < \beta$, and
\be
C_{2, \alpha,0}  ( |{\bf r}| ) =  
\frac{2}{\pi} \, \int^{\infty}_0 
\frac{ \lambda \, \sin (\lambda |{\bf r}|)}{ a_{\alpha} |\lambda|^{\alpha+2} + \varepsilon |\lambda|^2 } \, d \lambda .
\ee

This case is described by the fractional differential model introduced in Section 3 for the case of the order of Riesz fractional derivative is $\alpha+2>2$. 


To describe properties of electric field of the point charge
in the media with weak spatial dispersion, 
we consider properties of the function
\be
C_{2, \alpha,\beta}  (|{\bf r}|) =
\frac{2}{\pi} \, \int^{\infty}_0 
\frac{ \lambda \, \sin (\lambda |{\bf r}|)}{ 
a_{\beta} |\lambda|^{\beta+2} + 
a_{\alpha} |\lambda|^{\alpha+2}+ \varepsilon |\lambda|^2 } \, d \lambda ,
\ee
where $\beta> \alpha$.

Using the values for the sine integral $Si(x)$ for the infinite limit 
\be
\int^{\infty}_0 \frac{ \sin (z) }{z} = \frac{\pi}{2} ,
\ee
and the equation for the integral transform
(Section 2.3, equation (1) in \cite{BE}) of the form
\be
\int^{\infty}_0 z^{\alpha-1} \, \sin (z) = \Gamma(\alpha) 
\sin \left(\frac{\pi \beta}{2}\right) , \quad (-1 <\alpha <1) ,
\ee
we obtain the asymptotic for the function 
$C_{2, \alpha,\beta}  (|{\bf r}|) $ of the form
 
\be
C_{2, \alpha,\beta}  (|{\bf r}|) \approx
\frac{2}{\pi \varepsilon } \, \Bigl( \frac{\pi}{2} -
\frac{1}{|{\bf r}|^{\alpha}} \frac{a_{\alpha}}{\varepsilon} \, 
\Gamma(\alpha) \, \sin \left(\frac{\pi \alpha}{2}\right) 
- \frac{1}{|{\bf r}|^{\beta}} \frac{a_{\beta}}{\varepsilon} \, 
\Gamma(\beta) \, \sin \left(\frac{\pi \beta}{2}\right)\Bigr) .
\ee

It allows us to obtain the asymptotic behavior of 
the electrostatic potential 
\be \label{WSDP-1}
\Phi ({\bf r}) \ \approx \ 
\frac{Q}{4 \pi \varepsilon \varepsilon_0} \frac{1}{|{\bf r}|} -
\frac{a_{\alpha} Q }{4 \pi \varepsilon^2 \varepsilon_0} 
\frac{2 \Gamma(\alpha) \, \sin (\pi \alpha /2)}{\pi}
\frac{1}{|{\bf r}|^{\alpha+1}} -
\frac{a_{\beta} Q }{4 \pi \varepsilon^2 \varepsilon_0} 
\frac{2 \Gamma(\beta) \, \sin (\pi \beta /2)}{\pi}
\frac{1}{|{\bf r}|^{\beta+1}} ,
\ee
where $0< \alpha< \beta <1$.

The parameter $\varepsilon$ is interpreted as a relative permittivity of the media.
It is well known that 
far from the electric dipole the electrostatic potential of 
its electric field decreases with distance $|{\bf r}|$, 
as $|{\bf r}|^{-2}$ (see Section 40 in \cite{LL-2}), 
that is faster than the point charge potential ($\Phi ({\bf r})  \sim |{\bf r}|^{-1}$).

The first term in (\ref{WSDP-1}) describes the well-known Coulomb's field. 
The second and third terms in (\ref{WSDP-1}) look like changed dipole  
electrostatic field that for integer case ($\alpha=1$) has the from
$\Phi ({\bf r})  = d \cos \theta /
(4 \pi \varepsilon_0  \, |{\bf r}|^2 )$,  
where $d=|{\bf d}|$, and ${\bf d}$ is the (vector) dipole moment, 
and $\theta$ is an angle between the vectors
${\bf d}$ and ${\bf r}$ (see Section 40 in \cite{LL-2}).
We can consider the effective values
\be
d_{eff} (\alpha) = \frac{2 \Gamma (\alpha) a_{\alpha} Q }{\pi \varepsilon^2} , \quad 
\theta_{eff} = \frac{\pi}{2} (1-\alpha) 
\ee
for non-integer values of $\alpha$.
The second and third terms in equation (\ref{WSDP-1}) can be interpreted as a generalized 
dipole fields of power-law type with the non-integer orders $\alpha$ and $\beta$,
and these terms are represented as
\be \label{WSDP-1dip}
\Phi_{eff} ({\bf r}) =
- \frac{d_{eff}(\alpha) \cos \theta_{eff} (\alpha)}{4 \pi \varepsilon_0 \, |{\bf r}|^{\alpha+1}}  
- \frac{d_{eff}(\beta) \cos \theta_{eff} (\beta)}{4 \pi \varepsilon_0 \, |{\bf r}|^{\beta+1}} ,
\ee
where $0< \alpha< \beta <1$.

In Figure 4 present some plots (see Figure 3) of Coulomb's  electrostatic potentials $\Phi (|{\bf r}|) =(1/ 4 \pi \varepsilon_0) |{\bf r}|^{-1}$  and the potential with factor $C_{\beta,\alpha,2}(|{\bf r}|)$
for different orders of $0 < \alpha< \beta < 2$,
 where $a_{\beta}=a_{\alpha}=\varepsilon=1$. \\




\begin{figure}[H]
\begin{minipage}[h]{0.47\linewidth}
\resizebox{8cm}{!}{\includegraphics[angle=-90]{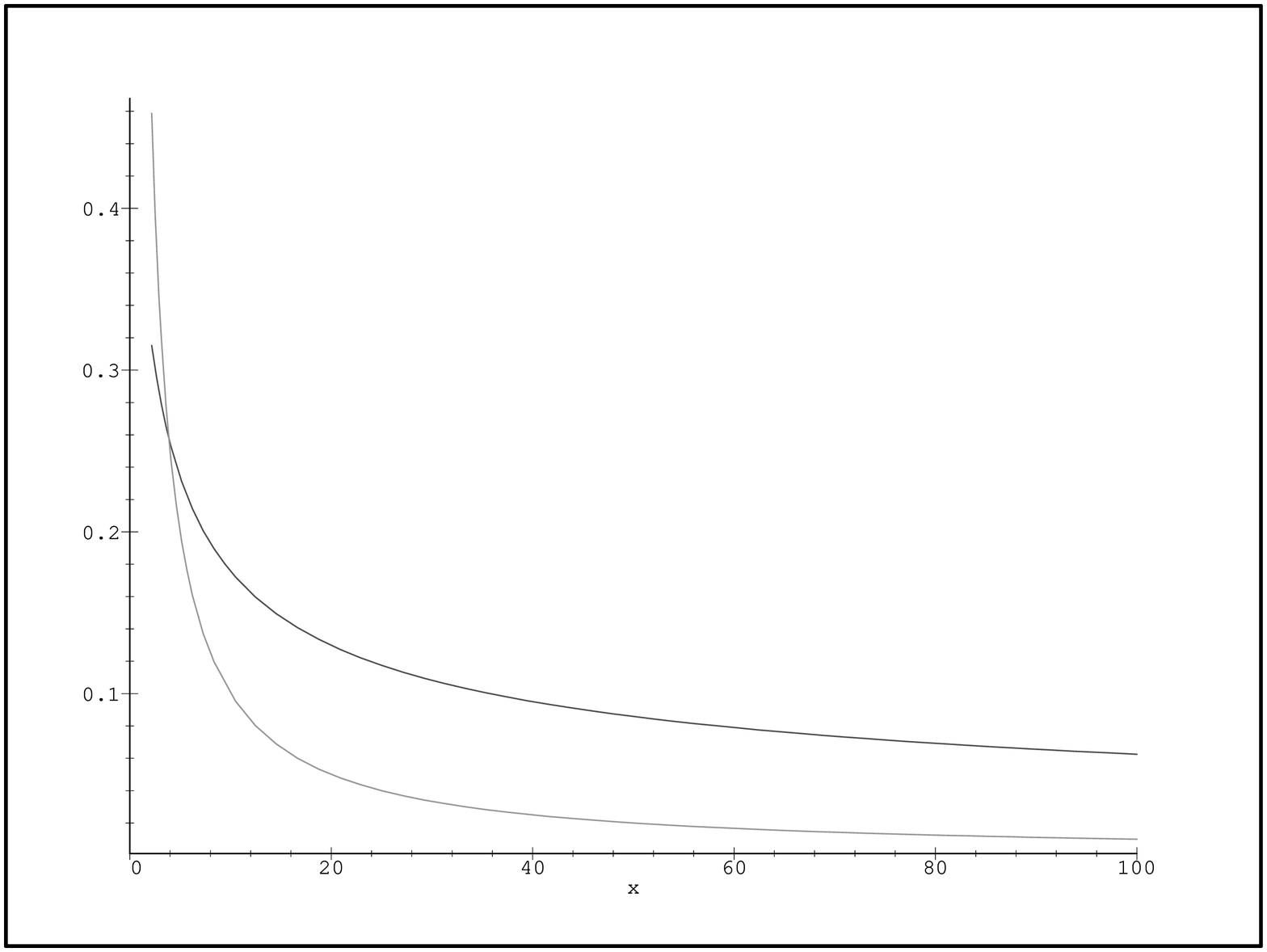}} 
a) \\
\end{minipage}
\hfill
\begin{minipage}[h]{0.47\linewidth}
\resizebox{8cm}{!}{\includegraphics[angle=-90]{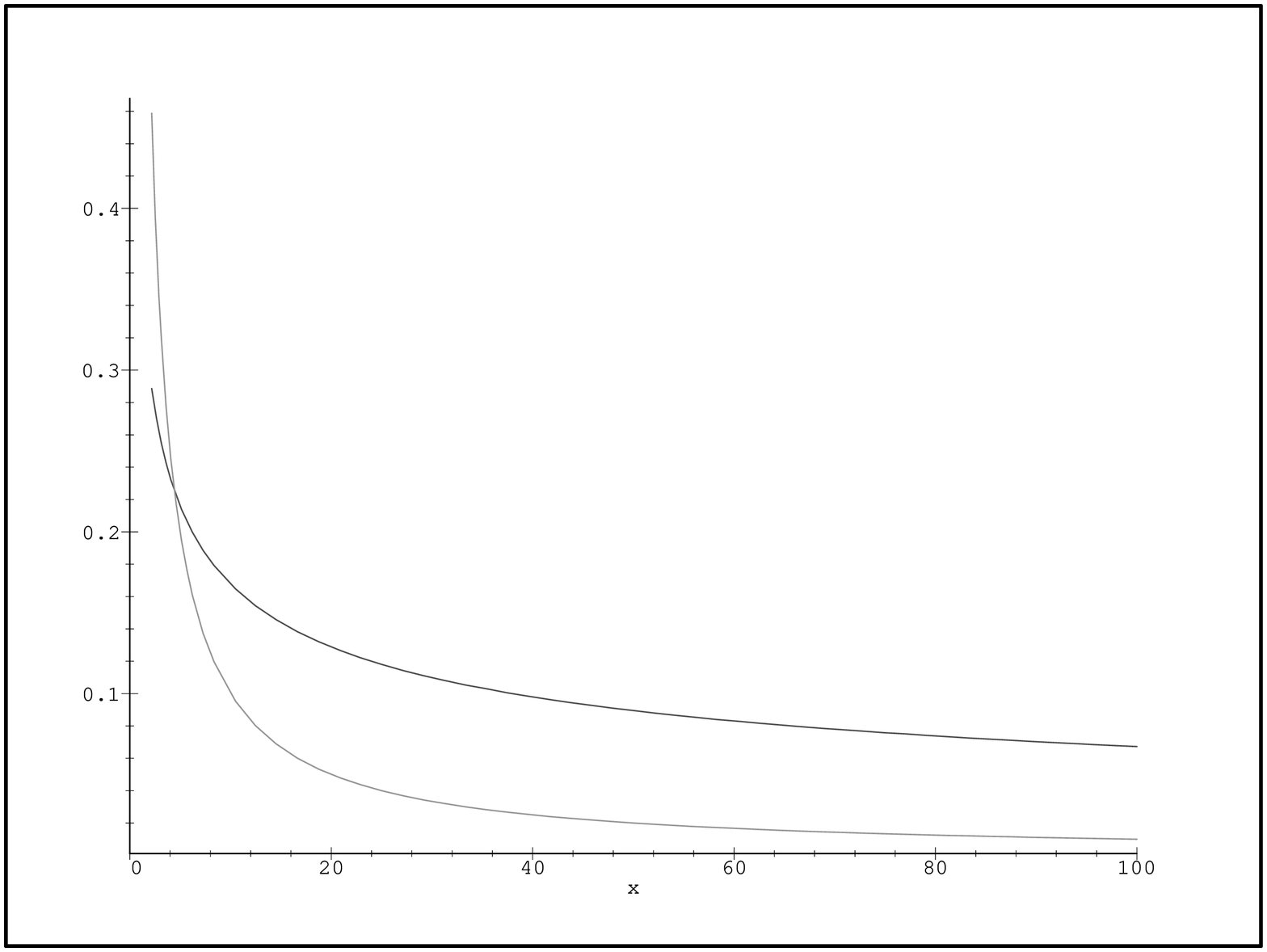}} 
 b) \\
\end{minipage}
\vfill
\begin{minipage}[h]{0.47\linewidth}
\resizebox{8cm}{!}{\includegraphics[angle=-90]{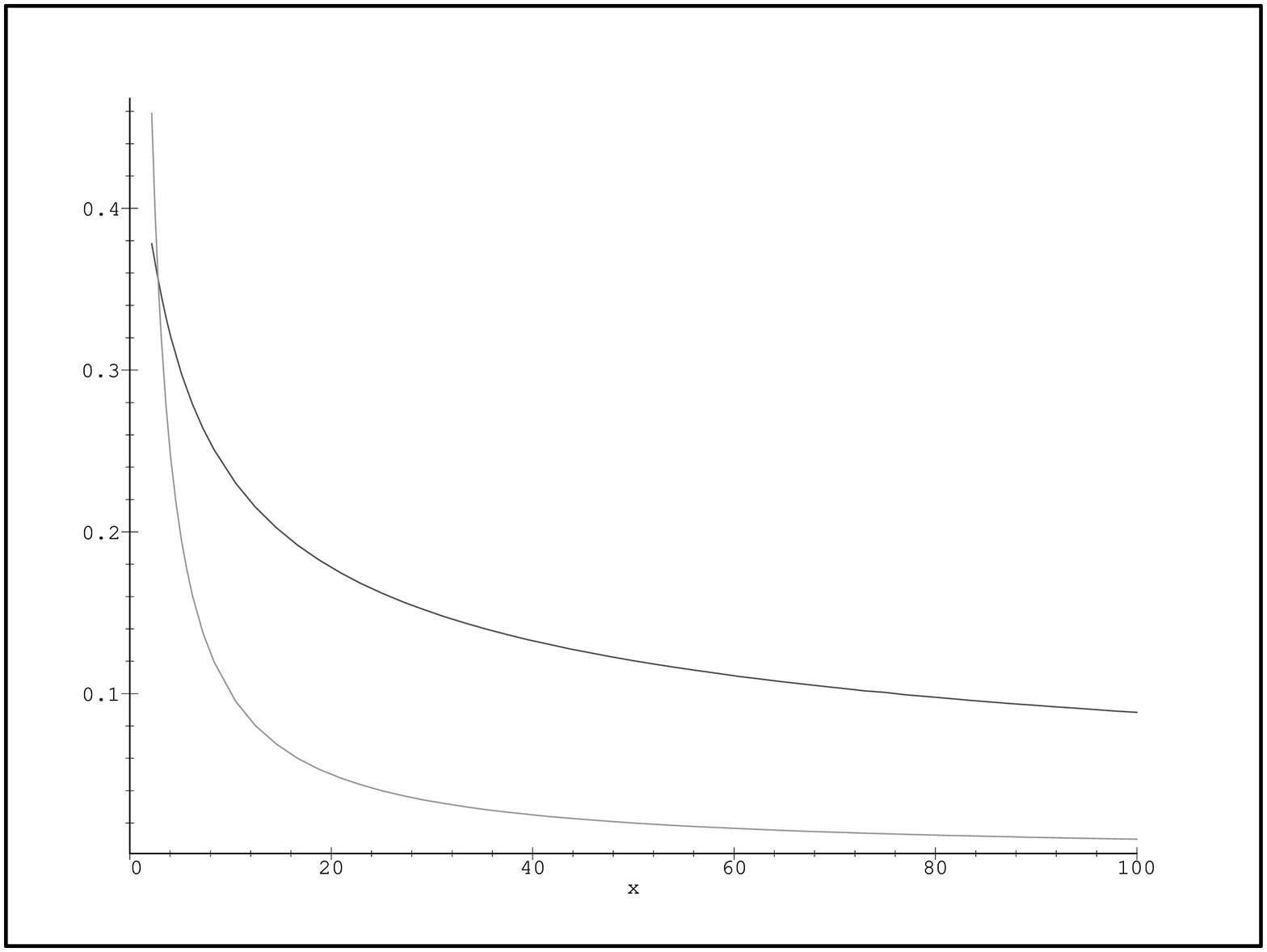}} 
c) \\
\end{minipage}
\hfill
\begin{minipage}[h]{0.47\linewidth}
\resizebox{8cm}{!}{\includegraphics[angle=-90]{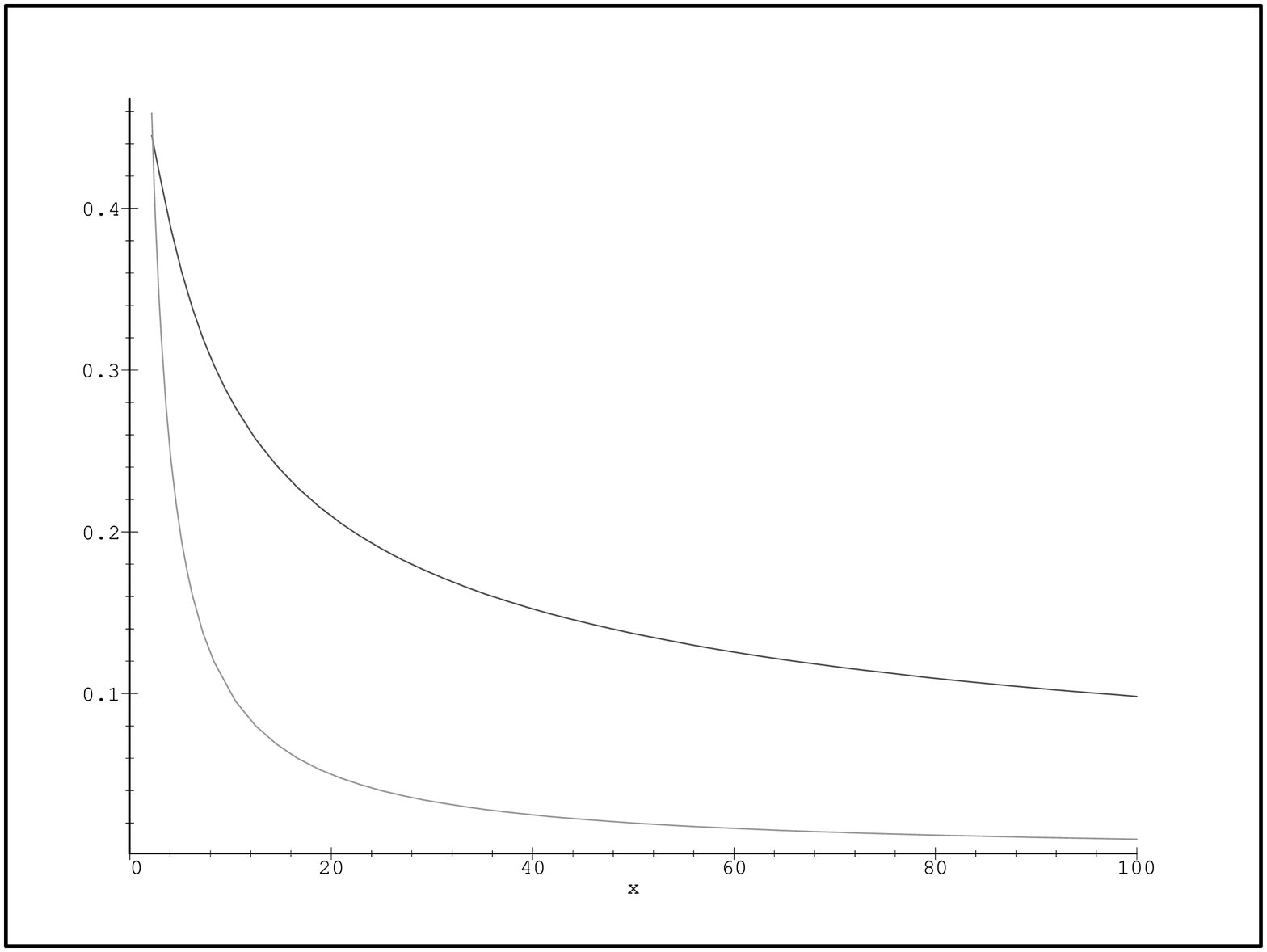}} 
d) \\
\end{minipage}
\caption{Plots of Coulomb's electrostatic potential $\Phi (x) =x^{-1}$ and 
the electrostatic potentials $y= x^{-1} \cdot C_{2,\alpha,\beta}(x)$ (the weak spatial dispersion case)
with $a_{\beta}=a_{\alpha}=\varepsilon=1$ 
for the orders: a) $\alpha=0.1$ and $\beta=1.1$, b)$\alpha=0.2$ and $\beta=0.4$, c) $\alpha=0.4$ and $\beta=1.4$, d) $\alpha=0.8$ and $\beta=1.8$. Here $x=|{\bf r}|$ and we use $0<x<100$. The asymptotic behavior for long distances is defined by $\beta$ only.}
\label{Plot4}
\end{figure}

From the plots (see Figure 4)
it is easy to see that far from the point charge in the media with weak
spatial dispersion the electrostatic potential 
decreases with distance $|{\bf r}|$ more slowly (\ref{WSDP-1})
than the potential of the point charge potential
($\Phi ({\bf r}) \sim |{\bf r}|^{-1}$).


\section{Conclusion}

We consider fractional power-law type generalizations of permittivity, 
and generalizations of the correspondent equations for electrostatic potential $\Phi ({\bf r})$ 
by involving the fractional generalization of the Laplacian \cite{Riesz-1,Riesz-2,SKM,KST}.
The simplest power-law forms of the longitudinal permittivity 
$\varepsilon_{\parallel} (|{\bf k}|) = \varepsilon_0 \Bigl( |{\bf k}|^{\alpha-2} + r^{-2}_D \, |{\bf k}|^{\beta-2} \Bigr)$
are suggested. 
The parameter $\alpha$ characterizes the deviation from Coulomb's law due to non-local properties of the medium.
The parameter $\beta$ characterizes the deviation from Debye's screening due to non-integer power-law type of non-locality in the medium.
The correspondent equation (\ref{Eq-3}) for electrostatic potential $\Phi ({\bf r})$ that has the form 
$((-\Delta)^{\alpha/2} \Phi) ({\bf r}) + r^{-2}_D ((-\Delta)^{\beta/2} \Phi)({\bf r}) = \varepsilon^{-1}_0 \rho({\bf r})$,
contains $(-\Delta)^{\alpha/2}$ and  $(-\Delta)^{\beta/2}$ are the Riesz fractional Laplacian 
\cite{Riesz-1,Riesz-2,SKM,KST}, and ${\bf r}$ and $r_D$ are dimensionless variables.

To the mentioned model can be find a explicit solution in terms of a Green type function. 
Also we will describe analytic solutions of the fractional differential equations (\ref{Eq-3}) for electrostatic potentials. 
The electrostatic potential of the point charge has form 
$\Phi ({\bf r}) = Q / (4 \pi \varepsilon_0 |{\bf r}|) \cdot C_{\alpha, \beta} (|{\bf r}|)$,
where $C_{\alpha, \beta} (|{\bf r}|)$ is defined by (\ref{Cab1}) describes the differences of Coulomb's potential and Debye's screening.
Using the analytic solutions of the fractional differential equations for electrostatic potentials,
we describe the asymptotic behaviors of the electrostatic potential.
The new type of behavior of the spatial-dispersion media with power-law non-locality is 
presented by power-law decreasing of the field at long distances instead of exponential decay.

In order to describe the properties of deviations separately, we consider the following special cases of the proposed model:

1) Fractional model of non-local deformation of Coulomb's law in the media with spatial dispersion that corresponds to the case $\beta=0$
and $\alpha \ne 2$, $\alpha >1$.
This model allows us to describe a possible deviation from Coulomb's law in the media with nonlocal properties defined by power-law type of spatial dispersion.

The electrostatic potential of the point charge in a media with this type of spatial dispersion has the form
$\Phi ({\bf r}) \ \sim \ |{\bf r}|^{\alpha-3}$ for $1<\alpha<2$ and $2<\alpha<3$ on small distances $|{\bf r}| \to 0$. 
In the case $\alpha>3$, we have the constant value of the potential for $|{\bf r}| \to 0$.
Therefore the electric field ${\bf E}$ of a point charge in the media with power-law type of 
spatial dispersion with $\alpha>3$ is equal to zero at small distances $|{\bf r}| \to 0$
that is analogous to the well-known case of the field inside a conducting charged 
sphere of the radius $R_{eff}$, for small distances.
The asymptotic behavior of potential for $|{\bf r}| \to \infty$ follow a power-law type also by our assumption. 
From the corresponding plots, we observe that the 
$C_{\alpha, 2}(|{\bf r}|)$ decreases more slowly than Debye's exponent $C_D(|{\bf r}|)$.
The function $C_{\alpha, 2}(|{\bf r}|)$ has a maximum for the case $2<\alpha<3$ and the maximum 
does not exists for $1<\alpha<2$, while for the particular case $\alpha=2$ it is well-known that it is the classical exponential Debye's screening. 

2) Fractional model of non-local deformation of Debye's screening in the media with spatial dispersion is defined by $\alpha=2$ and $0<\beta<2$.
Such model allows us to describe a possible deviation from  Debye's screening by non-local 
properties of the plasma-like media with the generalized power-law type of spatial dispersion. 

The generalized non-local properties deforms Debye's screening such 
that the exponential decay is replaced by the fractional power-law, 
and the electrostatic potential of the point charge in the media with this type of spatial dispersion is given by
$\Phi ({\bf r}) \ \sim \ |{\bf r}|^{\beta-3}$ for $0< \beta<2$ on the long distance $|{\bf r}| \to \infty$. 

3) Fractional non-local model that is described by equation (\ref{FPDE-4}) includes 
two parameters $(\alpha, \beta)$, where $\alpha>\beta > 0$ such that $\alpha \ne 2$.
In such model non-local properties deforms Coulomb's law and 
Debye's screening such that we have the fractional power-law decay  
for $0<\beta<3$ and $\alpha> \beta$. The asymptotic behavior $|{\bf r}| \to \infty$  does not depend on the parameter $\alpha$.
The field on the long distances is determined only by term with $(-\Delta)^{\beta/2}$ 
that can be interpreted  as a non-local deformation of Debye's term. 
It is remarkable that exist a maximum for the factor $C_{\alpha,\beta} (|{\bf r}|)$ in the case $0<\beta < 2 < \alpha$. 

The asymptotic behavior for $|{\bf r}| \to 0$  is given by
$\Phi ({\bf r}) \ \sim \ |{\bf r}|^{\alpha-3}$  for $1<\alpha<2$ and $2<\alpha<3$,
and $\Phi ({\bf r})$ is a constant for $\alpha>3$.
Note that the above asymptotic behavior does not depend on the parameters 
$\beta$ and $r^{-2}_D$.
The field on the short distances is determined only by term with $(-\Delta)^{\alpha/2}$, 
that can be considered as a non-local deformation of Coulomb's term. \\


We also consider weak spatial dispersion in the media with fractional power-law type of non-locality. 
In general, it cannot be describes by the usual Taylor approximation. 
The fractional Taylor series is very useful for approximating non-integer power-law functions. 
The media with spatial dispersion is described by the non-integer power-law type of functions $\varepsilon_{\parallel} (|{\bf k}|)$. 
Using fractional generalization of the Taylor series (see Appendix 2), we get
approximations of the form $\varepsilon_{\parallel}  (|{\bf k}|)/ \varepsilon_0  \approx \varepsilon + a_{\alpha} |{\bf k}|^{\alpha} + a_{\beta} |{\bf k}|^{\beta}$
for such type of media. 
If $\alpha=1$ and $\beta=2$, we have the usual Taylor formula, and
the well-known case of the weak spatial dispersion \cite{AG-1,AG-2,AG-3,AR-2,KR,LL-8,Halevi,UFN-1,UFN-2,UFN-3,UFN-4,UFN-5}.
In general, we should use a fractional generalization of the Taylor series. 
If the orders of the fractional Taylor series approximation 
will be correlated with the type of weak spatial dispersion, then the fractional Taylor series approximation 
for $\varepsilon_{\parallel} (|{\bf k}|)$ will be exact.

These fractional weak spatial dispersions is described by equation of the form
$a_{\beta} ( (-\Delta)^{(\beta+2)/2} \Phi)({\bf r}) + 
a_{\alpha} ( (-\Delta)^{(\alpha+2)/2} \Phi)({\bf r}) - 
\varepsilon \Delta \Phi ({\bf r}) = 
\frac{1}{ \varepsilon_0} \rho ({\bf r})$.
To this fractional differential equation 
we can be find a explicit solution in terms of a Green type function. 
Also we describe analytic solutions of the fractional differential equations for electrostatic potentials. 
It allows us to obtain the asymptotic behavior of the electrostatic potential of the form
$\Phi ({\bf r}) \ \approx \ \Phi_{Coulomb} ({\bf r}) + \Phi^{(\alpha)}_{eff, dipole} ({\bf r}) + \Phi^{(\beta)}_{eff, dipole} ({\bf r})$,
where $0< \alpha< \beta <1$.
The first term describes Coulomb's field. 
The second and third terms are interpreted as electrostatic fields generalized of changed dipoles of 
power-law type with the non-integer orders $\alpha$ and $\beta$, which have the form
$\Phi^{(\alpha)}_{eff} ({\bf r}) = - {d_{eff}(\alpha) \cos \theta_{eff} (\alpha)}/ (4 \pi \varepsilon_0 \, |{\bf r}|^{\alpha+1})$, 
where $0< \alpha< \beta <1$, $d_{eff}$ is the effective dipole moment, and $\theta$ is an effective angle.
For integer case ($\alpha=1$) we have the usual from $\Phi ({\bf r})  = d \cos \theta /(4 \pi \varepsilon_0  \, |{\bf r}|^2)$ of the fields.


\section*{Acknowledgments}

The first author thanks the Universidad de La Laguna for support and kind hospitality. This work was supported, in part, 
by Government of Spain grant No. MTM2010-16499 and by the President of Russian Federation grant for Science Schools No. 3920.2012.2.



\newpage
\section*{Appendix 1: Riesz fractional derivatives and integrals}

Let us consider Riesz fractional derivatives and fractional integrals.
The operations of fractional integration and fractional differentiation
in the $n$-dimensional Euclidean space $\mathbb{R}^n$
can be considered as fractional powers of the Laplace operator.
For $\alpha > 0$ and "sufficiently good" functions $f(x)$,
$x \in \mathbb{R}^n$, the Riesz fractional differentiation is defined \cite{Riesz-1,Riesz-2,SKM,KST} 
in terms of the Fourier transform ${\cal F}$ by
\be
(-\Delta)^{\alpha/2}_x f(x)= {\cal F}^{-1} \Bigl( |{\bf k}|^{\alpha} ({\cal F} f)({\bf k}) \Bigr) .
\ee
The Riesz fractional integration is defined by
\be
{\bf I}^{\alpha}_x f(x) =
{\cal F}^{-1} \Bigl( |{\bf k}|^{-\alpha} ({\cal F} f)({\bf k}) \Bigr) .
\ee

The Riesz fractional integration can be realized in the form of
the Riesz potential \cite{Riesz-1,Riesz-2,SKM,KST} defined as the Fourier convolution of the form
\be
{\bf I}^{\alpha}_x f(x)=
\int_{\mathbb{R}^n} K_{\alpha}(x-z) f(z) dz, \quad (\alpha >0) ,
\ee
where the function $K_{\alpha}(x)$ is the Riesz kernel.
If $\alpha>0$, and $\alpha \not=n,n+2,n+4,...$,
the function $K_{\alpha}(x)$ is defined by
\[ K_{\alpha}(x) = \gamma^{-1}_n(\alpha) |x|^{\alpha-n} . \]
If $\alpha \not=n,n+2,n+4,...$, then
\[ K_{\alpha}(x) = - \gamma^{-1}_n(\alpha) |x|^{\alpha-n} \ln |x| . \]
The constant $\gamma_n(\alpha)$ has the form
\be \gamma_n(\alpha)=
\begin{cases}
2^{\alpha} \pi^{n/2}\Gamma(\alpha/2)/ \Gamma(\frac{n-\alpha}{2}) &
\alpha \not=n+2j, \quad n \in \mathbb{N},
\cr
(-1)^{(n-\alpha)/2}2^{\alpha-1} \pi^{n/2} \;
\Gamma(\alpha/2) \;
\Gamma( 1+[\alpha-n]/2)
 & \alpha =n+2j.
\end{cases}
\ee

Obviously, the Fourier transform of the Riesz fractional integration is given by
\[ {\cal F} \Bigl( {\bf I}^{\alpha}_x f(x)\Bigr) =
|{\bf k}|^{-\alpha} ({\cal F} f)({\bf k}) . \]
This formula is true for functions $f(x)$ belonging to Lizorkin's space.
The Lizorkin spaces of test functions on $\mathbb{R}^n$
is a linear space of all complex-valued infinitely differentiable
functions $f(x)$ whose derivatives vanish at the origin:
\be
\Psi=\{ f(x): f(x) \in S(\mathbb{R}^n), \quad
(D^{\bf n}_xf)(0)=0, \quad |{\bf n}| \in \mathbb{N} \} ,
\ee
where $S(\mathbb{R}^n)$ is the Schwartz test-function space.
The Lizorkin space is invariant with respect
to the Riesz fractional integration.
Moreover, if $f(x)$ belongs to the Lizorkin space, then
\[ {\bf I}^{\alpha}_x f(x) {\bf I}^{\beta}_x f(x)
= {\bf I}^{\alpha+\beta}_x f(x) , \]
where $\alpha >0$, and $\beta >0$.

For $\alpha >0$, the Riesz fractional derivative
$(-\Delta)^{\alpha/2}=-\partial^{\alpha}/\partial |x|^{\alpha}$
can be defined in the form of the hypersingular integral (Sec. 26 in \cite{SKM}) by
\[ (-\Delta)^{\alpha/2}_x f(x)=\frac{1}{d_n(m,\alpha)} \int_{\mathbb{R}^n}
\frac{1}{|z|^{\alpha+n}} (\Delta^m_z f)(z) \, dz , \]
where $m> \alpha$, and $(\Delta^m_z f)(z)$ is a finite difference of
order $m$ of a function $f(x)$ with a vector step $z \in \mathbb{R}^n$
and centered at the point $x \in \mathbb{R}^n$:
\[ (\Delta^m_z f)(z) =\sum^m_{j=0} (-1)^j \frac{m!}{j! \, (m-j)!}  \, f(x-jz) . \]
The constant $d_n(m,\alpha)$ is defined by
\[ d_n(m,\alpha)=\frac{\pi^{1+n/2} A_m(\alpha)}{2^{\alpha}
\Gamma(1+\alpha/2) \Gamma(n/2+\alpha/2) \sin (\pi \alpha/2)} ,  \]
where
\[ A_m(\alpha)=\sum^m_{j=0} (-1)^{j-1} \frac{m!}{j!(m-j)!} \, j^{\alpha} . \]
Note that the hypersingular integral $ (-\Delta)^{\alpha/2}_x f(x)$ does not
depend on the choice of $m>\alpha$.

If $f(x)$ belongs to the space of "sufficiently good" functions, then
the Fourier transform ${\cal F}$ of the Riesz fractional derivative
is given by
\[ ({\cal F} (-\Delta)^{\alpha/2} f)({\bf k}) = |{\bf k}|^{\alpha} ({\cal F}f)({\bf k}) . \]
This equation is valid for the Lizorkin space \cite{SKM}
and the space $C^{\infty}(\mathbb{R}^n)$ of infinitely differentiable
functions on $\mathbb{R}^n$ with compact support.

The Riesz fractional derivative yields an operator inverse
to the Riesz fractional integration for a special space of functions.
The formula
\be \label{bfDa}
(-\Delta)^{\alpha/2}_x \, {\bf I}^{\alpha}_x f(x) = f(x) , \quad (\alpha >0) \ee
holds for "sufficiently good" functions $f(x)$.
In particular, equation (\ref{bfDa}) for $f(x)$ belonging to the Lizorkin space.
Moreover, this property is also valid for the Riesz fractional integration in
the frame of $L_p$-spaces: $f(x) \in L_p(\mathbb{R})$ for$1 \leqslant p < n/a$.
Here the Riesz fractional derivative $(-\Delta)^{\alpha/2}_x$
is understood to be conditionally convergent in the sense that
\be \label{bfDae}
(-\Delta)^{\alpha/2}_x =
\lim_{\epsilon \to  0} (-\Delta)^{\alpha/2}_{x, \epsilon} , \ee
where the limit is taken in the norm of the space $L_p(\mathbb{R})$, and
the operator $(-\Delta)^{\alpha/2}_{x, \epsilon}$ is defined by
\[ (-\Delta)^{\alpha/2}_{x, \epsilon}=
\frac{1}{d_n(m,\alpha)} \int_{|z| > \epsilon}
\frac{1}{|z|^{\alpha+n}} (\Delta^m_z f)(z) \, dz , \]
where $m> \alpha$, and $(\Delta^m_z f)(z)$ is a finite difference of
order $m$ of a function $f(x)$ with a vector step $z \in \mathbb{R}^n$
and centered at the point $x \in \mathbb{R}^n$.
As a result, the following property holds.
If $0<\alpha <n$ and $f(x) \in L_p(\mathbb{R})$ for$1 \leqslant p < n/a$, then
\[ (-\Delta)^{\alpha/2}_x \, {\bf I}^{\alpha}_x f(x) = f(x) , \quad (\alpha >0) , \]
where $(-\Delta)^{\alpha/2}_x$  is understood in the sense of (\ref{bfDae}),
with the limit being taken in the norm of the space $L_p(\mathbb{R})$.
This result is proved in \cite{SKM} (see Theorem 26.3).

We note that the Riesz fractional derivatives appear in
the continuous limit of lattice models with long-range interactions \cite{PH7}.

\section*{Appendix 2: Fractional Taylor Formula}

\subsection*{Riemann-Liouville and Caputo derivatives}

The left-sided Riemann-Liouville derivatives of order $\alpha >0$ are defined by
\be
(\, ^{RL}D^{\alpha}_{a+} f)(x) = \frac{1}{\Gamma(n-\alpha)} \left( \frac{d}{dx}\right)^n 
\int^x_a \frac{f(x^{\prime}) \, dx^{\prime}}{(x-x^{\prime})^{\alpha-n+1}} ,
\quad (n=[\alpha]+1) .
\ee
We can rewrite this relation in the form
\be
(\, ^{RL}D^{\alpha}_{a+} f)(x) =\left( \frac{d}{dx}\right)^n \, (I^{n-\alpha}_{a+} f)(x) ,
\ee
where $I^{\alpha}_{a+}$ is a  left-sided Riemann-Liouville integral of order $\alpha >0$ 
\be \label{RLI}
(I^{\alpha}_{a+} f)(x) = \frac{1}{\Gamma(\alpha)} 
\int^x_a \frac{f(x^{\prime}) \, dx^{\prime}}{(x-x^{\prime})^{1-\alpha}} ,
\quad (x>a) .
\ee

The Caputo fractional derivative of order $\alpha$ is defined by
\be
(\, ^CD^{\alpha}_{a+} f)(x) = \left( I^{n-\alpha}_{a+} \left( \frac{d}{dx}\right)^n f \right)(x) ,
\ee
where $I^{\alpha}_{a+}$ is a  left-sided Riemann-Liouville integral (\ref{RLI}) of order $\alpha >0$.
In equation (\ref{Tay-Cap}) we use $0<\alpha<1$ and $n=1$. 
The main distinguishing feature of the Caputo fractional derivative is that, 
like the integer order derivative, 
the Caputo fractional derivative of a constant is zero. 

Note also that the third term in (\ref{Tay-Cap}) involves the fractional derivative 
of the fractional derivative, which is not the same as the $2 \alpha$ fractional
derivative. In general,
\[ (\, ^CD^{\alpha}_{a+} \, ^CD^{\alpha}_{a+} f)(x) \ne (\, ^CD^{2\alpha}_{a+} f)(x) . \]
Then the coefficients of the fractional Taylor
series can be found in the usual way, by repeated differentiation.
This is to ensure that the fractional derivative of order $\alpha$ 
of the function $(x-a)^{\alpha}$ is a constant. 
The repeated the fractional derivative of order $\alpha$ gives zero. 
Then the coefficients of the fractional Taylor
series can be found in the usual way, by repeated differentiation.

\subsection*{Fractional Taylor series in the Riemann-Liouville form}

Let $f(x)$ be a real-valued function such that 
the derivative $(\, ^{RL}D^{\alpha+m}_{a+} f)(x)$ is integrable. 
Then the following analog of Taylor formula holds 
(see Chapter 1. Section 2.6 \cite{SKM}):
\be
f(x)= \sum^{m-1}_{j=0} \frac{(\, ^{RL}D^{\alpha+j}_{a+}f)(a+)}{\Gamma(\alpha+j+1)} \, 
(x-a)^{\alpha+j} +R_m (x) , \quad (\alpha>0) ,
\ee
where $D^{\alpha+j}_{a+}$ are left-sided Riemann-Liouville derivatives, and
\be
R_m (x) = (I^{\alpha+m}_{a+} \, ^{RL}D^{\alpha+m}_{a+} f)(x) .
\ee

\subsection*{Riemann formal version of the generalized Taylor series}

The Riemann formal version of the generalized Taylor series \cite{Riem,Hardy}:
\be
f(x)= \sum^{+\infty}_{m=-\infty} \frac{(\, ^{RL}D^{\alpha+m}_a f)(x_0)}{\Gamma(\alpha+m+1)} (x-x_0)^{\alpha+m} ,
\ee
where $\, ^{RL}D^{\alpha}_a$ for $\alpha>0$ is the Riemann-Liouville fractional derivative, and
$\, ^{RL}D^{\alpha}_a=I^{-\alpha}_a$ for $\alpha<0$ is the Riemann-Liouville fractional integral of order $|\alpha|$.

\subsection*{Fractional Taylor series in the Trujillo-Rivero-Bonilla form}

The Trujillo-Rivero-Bonilla form of the generalized Taylor formula \cite{TRB}:
\be
f(x) = \sum^{m}_{j=0} \frac{c_j}{\Gamma((j+1)\alpha)} \, (x- a)^{(j+1)\alpha-1}  +R_m(x,a),
\ee
where $\alpha \in [0;1]$, and
\be
c_j = \Gamma(\alpha)  \, [(x-a)^{1-\alpha} \, (\, ^{RL}D^\alpha_a)^j f(x)](a+) ,
\ee
\be
R_m(x,a) = \frac{ ((\, ^{RL}D^{\alpha}_a)^{m+1} f)(\xi) }{ \Gamma((m+1) \alpha+1)} \,
 (x-a)^{(m+1)\alpha} , \quad \xi \in [a;x] .
\ee

\subsection*{Fractional Taylor series in the Dzherbashyan-Nersesian form}

Let $\alpha_k$, $(k=0,1,...,m)$ be increasing sequence of real numbers such that
\be
0 < \alpha_k-\alpha_{k-1} \le 1, \quad \alpha_0=0, \quad k=1,2,...,m.
\ee

We introduce the notation \cite{Arm1,Arm2} (see also Section 2.8 in \cite{SKM}):
\be
D^{(\alpha_k)}  = I^{1-(\alpha_k-\alpha_{k-1} )}_{0+} D^{1+ \alpha_{k-1} }_{0+} .
\ee
In general, $D^{(\alpha_k)} \ne ^{RL}D^{\alpha_k}_{0+} $. 
Fractional derivative  $D^{(\alpha_k)}$ differs from the Riemann-Liouville derivative 
$^{RL}D^{\alpha_k}_{0+} $ by finite sum of power functions since (see Eq. 2.68 in \cite{KST})
\be
I^{\alpha}_{0+} I^{\beta}_{0+} \ne I^{\alpha+\beta}_{0+} .
\ee
 
The generalized Taylor formula \cite{Arm1,Arm2}
\be
f(x) = \sum^{m-1}_{k=0} a_k \, x^{\alpha_k} +R_{m}(x), \quad (x>0).
\ee
where
\be
a_k = \frac{(D^{(\alpha_k)}  f)(0)}{\Gamma(\alpha_k+1)} , \quad
R_{m}(x) = \frac{1}{\Gamma(\alpha_m+1)} \int^x_0 (x-z)^{\alpha_m-1} \, (D^{(\alpha_k)}  f)(z) \, dz .
\ee

\subsection*{Fractional Taylor series in the Odibat-Shawagfeh form}

The fractional Taylor series is a generalization of the Taylor series 
for fractional derivatives, where $\alpha$ is the fractional
order of differentiation, $0<\alpha<1$. 
The fractional Taylor series with Caputo derivatives \cite{OdbSh}
has the form
\be \label{Tay-Cap}
f(x)=f(a)+ \frac{(\, ^CD^{\alpha}_{a+}f)(a)}{\Gamma (\alpha+1)} (x-a)^{\alpha}
+  \frac{(\, ^CD^{\alpha}_{a+} \, ^CD^{\alpha}_{a+} f)(a)}{\Gamma (2\alpha+1)} (x-a)^{2 \alpha}+ . . . ,
\ee
where $\, ^CD^{\alpha}_{a+}$ is the Caputo fractional derivative of order $\alpha$.


\end{document}